\documentclass[
 reprint,
superscriptaddress,
 amsmath,amssymb,
 aps,
]{revtex4-2}

\usepackage{graphicx}
\usepackage{dcolumn}
\usepackage{bm}
\usepackage{float}
\usepackage{epsfig}
\usepackage{amsfonts}
\usepackage{amssymb}
\usepackage{amsmath}
\usepackage{multirow}
\newcommand{\lcdm}{$\Lambda$CDM}
\usepackage{rsfso}
\usepackage{mathrsfs}
\usepackage{subcaption}
\usepackage[justification=justified, format=plain]{caption}
\usepackage{booktabs}
\usepackage[normalem]{ulem}
\usepackage{placeins}
\usepackage{xcolor}
\DeclareMathAlphabet{\mathcal}{OMS}{cmsy}{m}{n}
\usepackage[hidelinks]{hyperref}
\hypersetup{colorlinks=false,breaklinks=true}

\def\sqdeg{deg$^2$}
\def\uk{\ensuremath{\mu \mathrm{K}}}
\def\ukarcmin{\uk-{\rm arcmin}}

\newcommand{\sptthreeg}{SPT-3G}

\def\sptnew{SPT-3G}
\newcommand{\mainfield}{\mbox{\sptthreeg{} Main-1500}}
\newcommand{\summerfield}{\mbox{\sptthreeg{} Summer}}
\newcommand{\sptsummerwinter}{\mbox{\sptthreeg{} Ext-4k}}
\newcommand{\sptextfield}{\mbox{\sptthreeg{} Ext-10k}}

\newcommand{\mnras}{MNRAS}
\newcommand{\aap}{A\&A}
\newcommand{\apjl}{Astrophys. J. Lett.}

\newcommand{\fsky}{f_\mathrm{sky}}
\newcommand{\clcov}{\mathbf{C}_\mathrm{\ell}}
\newcommand{\clinv}{\clcov^{-1}}

\begin{document}

\title{Cool dark sector, concordance, and a low {\boldmath$\sigma_8$}}

\author{Ellie Hughes}
 \email{ellieh@mit.edu}
 \affiliation{Department of Physics, Bryn Mawr College, Bryn Mawr, Pennsylvania 19010, USA}
  \affiliation{Department of Physics and Astronomy, University of California, Davis, California 95616, USA}
  \affiliation{Departamento de Física, Facultad de Ciencias Físicas y Matemáticas, Universidad de Chile, Santiago, Chile}
\author{Fei Ge}
 \email{fge@ucdavis.edu}
 \affiliation{Department of Physics and Astronomy, University of California, Davis, California 95616, USA}
\author{Francis-Yan Cyr-Racine}
 \affiliation{Department of Physics and Astronomy, University of New Mexico, Albuquerque, New Mexico 87106, USA}
\author{Lloyd Knox}
 \affiliation{Department of Physics and Astronomy, University of California, Davis, California 95616, USA}
\author{Srinivasan Raghunathan}
 \affiliation{Center for AstroPhysical Surveys, National Center for Supercomputing Applications, Urbana, Illinois 61801, USA}

\date{May 10, 2024}

\begin{abstract}
We investigate a cosmological model in which a fraction of the dark matter is atomic dark matter (ADM). This ADM consists of dark versions of the electron and of the proton, interacting with each other and with dark photons just as their light sector versions do, but interacting with everything else only gravitationally. We find constraints given current cosmic microwave background (CMB) and baryon acoustic oscillation (BAO) data, with and without an $H_0$ prior, and with and without enforcing a big bang nucleosynthesis consistent helium abundance. We find that, at low dark photon temperature, one can have consistency with BAO and CMB data, with a fraction of dark matter that is ADM ($f_{\rm adm}$) as large as $\sim 0.1$. Such a large $f_{\rm adm}$ leads to a suppression of density fluctuations today on scales below about 60 Mpc that may be of relevance to the $\sigma_8$ tension. Our work motivates calculation of nonlinear corrections to matter power spectrum predictions in the ADM model. We forecast parameter constraints to come from future ground-based CMB surveys, and find that if ADM is indeed the cause of the $\sigma_8$ tension, the influence of the ADM, primarily on CMB lensing, will likely be detectable at high significance. 

\end{abstract}

\maketitle

\section{Introduction} \label{sec:introduction}

The physics of dark matter remains a mystery. While weakly interacting massive particles (WIMPs) \cite{goodman1985,jungman1996} and the quantum chromodynamic axion \cite{peccei1977,wilczek1978} remain leading dark matter candidates, the possibility that dark matter could be part of an extended ``dark sector'' has gained traction in the last decade \cite{arkanihamed2009}. One intriguing possibility is that this dark sector could contain a copy of the Standard Model field content, such as in the twin Higgs model (see, e.g., Refs.~\cite{Chacko:2005pe,Chacko:2005vw,Chacko:2005un,Barbieri:2005ri,Craig:2013fga,Craig:2015pha,GarciaGarcia:2015fol,Craig:2015xla,Farina:2015uea,Farina:2016ndq,Prilepina:2016rlq,Barbieri:2016zxn,Craig:2016lyx,Berger:2016vxi,Chacko:2016hvu,Csaki:2017spo,Chacko:2018vss,Elor:2018xku,Hochberg:2018vdo,Francis:2018xjd,Harigaya:2019shz,Ibe:2019ena,Dunsky:2019upk,Csaki:2019qgb,Koren:2019iuv,Terning:2019hgj,Johns:2020rtp,Roux:2020wkp,Ritter:2021hgu,Curtin:2021alk,Curtin:2021spx}). In such scenarios, part of the dark matter can be charged under a new $U_{\rm d}(1)$ gauge interaction \cite{Blinnikov:1983gh,Ackerman:2008gi,Feng:2009mn,Agrawal:2016quu,Foot:2002iy,Foot:2003jt,Foot:2004pa,Foot:2004wz,Foot:2007iy,Foot:2011ve,Foot:2013vna,Foot:2014mia,Foot:2014uba,Foot:2016wvj,Ciarcelluti:2004ik,Ciarcelluti:2004ip,Ciarcelluti:2008vs,Ciarcelluti:2010zz,Ciarcelluti:2012zz,Ciarcelluti:2014scd,Cudell:2014wca} and even form atom-like bound states \cite{Goldberg:1986nk,Fargion:2005ep,Khlopov:2005ew,Khlopov:2008ty,Kaplan:2009de,Khlopov:2010pq,Kaplan:2011yj,Khlopov:2011tn,Behbahani:2010xa,Cline:2012is,cyr-racine_cosmology_2013,Cline:2013pca,Cyr-Racine:2013fsa,Fan:2013tia,Fan:2013yva,McCullough:2013jma,Randall:2014kta,Khlopov:2014bia,Pearce:2015zca,Choquette:2015mca,Petraki:2014uza,Cirelli:2016rnw,Petraki:2016cnz,Curtin:2020tkm}. In its simplest incarnation, this ``atomic'' dark matter (ADM) involves two massive dark fermions (a ``dark proton'' and a ``dark electron'') interacting with a massless dark photon. In the early Universe, such a dark sector forms an ionized plasma in which the dark fermions are tightly coupled with the dark photon bath. Once its temperature falls below the binding energy between the two dark fermions, neutral dark atoms can form in a process similar to standard hydrogen recombination \cite{Peebles:1968ja}. 

This ADM scenario can have an important impact on the way structure assembles in our Universe. At early times, radiation pressure from the $U_{\rm d}(1)$ dark photon bath opposes gravitational infall of dark matter into potential wells, hence slowing down the growth of structure and modifying the anisotropies of the cosmic microwave background (CMB). At late times, dark atoms behave as a form of dissipative dark matter, allowing it to cool and form structures such as dark disks \cite{Fan:2013tia,Fan:2013yva,Randall:2014kta,Schutz:2017tfp,Ghalsasi:2017jna,Widmark:2021gqx,Roy:2023zar}, mirror stars \cite{Curtin:2019lhm,Curtin:2019ngc}, and exotic compact objects \cite{Shandera:2018xkn,Singh:2020wiq,Ryan:2021dis,Ryan:2022hku,Gurian:2021qhk,Gurian:2022nbx,Fernandez:2022zmc}.

Given its possible impact on a broad range of scales, it is natural to ask if ADM, whether it forms part or the entirety of the dark matter, could be playing a role in some of the tensions facing the $\Lambda$ Cold Dark Matter ($\Lambda$CDM) cosmological model \cite{divalentino21}. A particularly relevant one is the so-called $\sigma_8$ tension, which refers to a potential discrepancy between different measurements of the matter fluctuation amplitude on $8h^{-1}\,\textrm{Mpc}$ scales \cite{di_valentino_cosmology_2021}. Assuming standard $\Lambda$CDM cosmology, CMB observations from \emph{Planck} find $\sigma_8 = 0.811 \pm 0.006$ \cite{planck_collaboration_planck_2020}. However, measurements of cosmic shear from the KiDS-1000 survey find $\sigma_8 = 0.76^{+0.025}_{-0.020}$ \cite{heymans_kids-1000_2021}. Although these two specific measurements are only in $\sim 3\sigma$ disagreement, multiple observations of the large-scale structure of the Universe using a variety of techniques have found $\sigma_8$ values that are consistently lower than those inferred from the CMB \cite{HSC:2018mrq, heymans_kids-1000_2021, Hang:2020gwn, Garcia-Garcia:2021unp, DES:2021wwk, White:2021yvw, DES:2022xxr}, potentially pointing to an underlying problem with $\Lambda$CDM as the standard model of cosmology.

The cosmological impact of ADM on the large-scale structure of the Universe has been extensively studied in the literature (see, e.g., Refs.~\cite{Kaplan:2009de,Kaplan:2011yj,cyr-racine_cosmology_2013,Cyr-Racine:2013fsa,Chacko:2018vss,Bansal:2021dfh,bansal_precision_2022}). Much like in the visible sector where the baryon acoustic oscillations (BAO) get imprinted on the matter power spectrum, the large dark photon radiation pressure at early times can leave a dark acoustic oscillation (DAO) pattern on the spectrum of matter fluctuations. On scales smaller than this ``dark'' acoustic horizon, such a DAO pattern is typically damped due to the finite mean free path of dark photons, in a process reminiscent of Silk damping \cite{Silk:1967kq}. However, if the dark sector is significantly cooler than the visible sector, the relatively weak pressure support (which scales as the fourth power of the dark sector temperature) leads to a heavily damped DAO spectrum. In this case, the main impact of ADM on structure formation is a suppressed matter power spectrum on scales smaller than the dark acoustic horizon. This suppression, which also occurs in other models involving dark matter-dark radiation interaction (see, e.g., Refs.~\cite{Buen-Abad:2015ova,Chacko:2016kgg,Buen-Abad:2017gxg,Joseph:2022jsf,Buen-Abad:2022kgf,Rubira:2022xhb,Buen-Abad:2023uva,Aloni:2021eaq,Gariazzo:2023hch,Schoneberg:2023rnx}), could play an important role in addressing the $\sigma_8$ tension. 

In this paper, we study the impact that ADM could have on the $\sigma_8$ tension while maintaining concordance with a broad range of cosmological data including CMB, BAO, Cepheid-calibrated supernovae constraints on the Hubble constant, and the primordial abundance of helium. We show that a cool dark sector made of dark atoms could offer a possible solution to the $\sigma_8$ tension. We also present forecasts for how future observations could help constrain this scenario. 

This paper is organized as follows. In Sec.~\ref{sec:modelspace}, we outline the parameter space explored and the physics of the ADM model. In Sec.~\ref{sec:constraints}, we present the constraints placed by current data on ADM parameters, including an allowed region of low $\sigma_8$ and low dark photon temperature. In Sec.~\ref{sec:lowsigma8}, we confirm the observational viability of this allowed region and explore its impacts on matter power spectra and CMB power spectra. In Sec.~\ref{sec:forecasts}, we forecast the sensitivity of future CMB experiments to ADM in the low $\sigma_8$ and low dark photon temperature regime. In Sec.~\ref{sec:conclusions}, we summarize and conclude.

\section{Physical effects of changing atomic dark matter parameters} \label{sec:modelspace}

We examine in this section the impact ADM has on CMB and matter power spectra, and how those impacts depend on ADM parameters. Because of the possibility for resolving the $\sigma_8$ tension, we focus on the ``cool" dark sector regime, as mentioned in the introduction, where the dark photon temperature today $T_{{\rm d}\gamma}^0$ is between 0.6 K and 1.0 K. Prior work by some of us \cite{ge_scaling_2022} identified the fraction of non-relativistic matter that is pressure supported as a useful quantity for understanding changes in observables. We will indeed find that to be the case here.

Before focusing on the cool regime, we begin with a quick overview of the ADM model and its phenomenology. More detailed discussions can be found in Refs.~\cite{Goldberg:1986nk,Fargion:2005ep,Khlopov:2005ew,Khlopov:2008ty,Kaplan:2009de,Khlopov:2010pq,Kaplan:2011yj,Khlopov:2011tn,Behbahani:2010xa,Cline:2012is,cyr-racine_cosmology_2013,Cline:2013pca,Cyr-Racine:2013fsa,Fan:2013tia,Fan:2013yva,McCullough:2013jma,Randall:2014kta,Khlopov:2014bia,Pearce:2015zca,Choquette:2015mca,Petraki:2014uza,Cirelli:2016rnw,Petraki:2016cnz,Francis:2018xjd,Curtin:2020tkm}. The ADM consists of massive dark protons and dark electrons interacting with massless dark photons. At early times, dark sector particles are highly coupled together, forming a dark plasma. Via a Thompson-scattering-like interaction, dark electrons scatter off dark photons at a per-particle rate higher than the cosmic expansion rate. When the dark photons cool down, dark protons and dark electrons can form bound states, similar to the recombination process in the visible sector. After decoupling from the dark baryons, dark photons freely stream with a mean free path larger than the Hubble distance. Dark baryons, after dark recombination, experience very little pressure support and so evolve much like cold dark matter (CDM). 

There are five parameters to characterize the atomic dark sector: the density of dark baryons $\omega_{\rm db}$, the temperature of the dark photons today $T_{{\rm d}\gamma}^0$, the binding energy of dark hydrogen $B_{\rm d}$, the mass of dark hydrogen $m_{\rm d}$, and the dark fine structure constant $\alpha_{\rm d}$. However, at fixed $B_{\rm d}/T_{{\rm d}\gamma}^0$ and $\omega_{\rm db}$, we expect that varying $m_{\rm d}$ and $\alpha_{\rm d}$ would have very little impact on the observables we consider, so for simplicity we fix them to their light sector values. In contrast, the ratio $B_{\rm d}/T_{{\rm d}\gamma}^0$ has a significant impact on observables via its influence on the redshift of dark recombination.

In previous work \cite{cyr-racine_symmetry_2022, ge_scaling_2022}, the ADM model was used to mimic a scaling transformation that preserves dimensionless cosmological observables. To do so, both the dark hydrogen binding energy to dark photon temperature ratio ($B_{\rm d}/T_{{\rm d}\gamma}^0$) and the dark baryon-to-photon number ratio ($\eta_{\rm d}$) are set to their values in the visible sector. The same $B_{\rm d}/T_{{\rm d}\gamma}^0$ ensures recombination happens around the same time in both dark and visible sectors. The same baryon-to-photon ratio ensures that the photon and baryon perturbations are in phase in both sectors. In this work, we drop these requirements on the values of $B_{\rm d}/T_{{\rm d}\gamma}^0$ and $\eta_{\rm d}$, allowing them to vary freely. As a result, the redshift of dark recombination is free to vary, as is the dark photon temperature. This section is aimed at elucidating the impact on power spectra of variation of these two parameters.

Let us specify our fiducial \lcdm\ and ADM models. The fiducial \lcdm\ model is the \lcdm\ model that best fits the \emph{Planck} 2018 TT,TE,EE+lowE+lensing \cite{planck_collaboration_planck_2020-1} likelihoods. We refer to it as the base model. It has parameter values, 
\begin{align} \label{eq:basemodelpars}
    &\{ H_0, \Omega_{\rm b} h^2,\Omega_{\rm dm}h^2, \tau, A_{\rm s},  n_{\rm s}, Y_{\rm P}, N_{\rm eff}, \Sigma m_\nu\} \\
    =& \{{67.37\ {\rm  km/s/Mpc }}, 0.02233, 0.1198, 0.0540, {2.097\times10^{-9}}, \nonumber \\ & 0.9652, 0.2454, 3.046, 0.06\ {\rm eV}\}.\nonumber
\end{align}
For the base model, the dark matter consists only of CDM, namely $\Omega_{\rm dm}h^2 = \Omega_{\rm c}h^2$. For ADM models, we use three additional parameters to specify the dark sector properties: $T_{{\rm d}\gamma}^0$, $B_{\rm d}/T_{{\rm d}\gamma}^0$, and the fraction of dark matter in dark baryons, 
\begin{equation}
    f_{\rm adm} \equiv\frac{\omega_{\rm db}}{\omega_{\rm dm}} = \frac{\omega_{\rm db}}{\omega_{\rm db} + \omega_{\rm c}}.
\end{equation}
For the fiducial ADM model, we fix the total dark matter density to be the same as in the base model, 
\begin{equation}
\omega_{\rm dm}^{\rm ADM} = \omega_{\rm dm}^{\rm \Lambda CDM},
\label{eq:fidadm1}
\end{equation}
fix other cosmological parameters to the values in Eq.~\eqref{eq:basemodelpars}, and pick the following ADM parameters:
\begin{equation}
    \{f_{\rm adm}, B_{\rm d}/T_{{\rm d}\gamma}^0, T_{{\rm d}\gamma}^0\} = \{0.1, \ 10 \ \mathrm{eV/K}, \ 0.6\ \mathrm{K}\}. \label{eq:fidadm2}
\end{equation}
By setting $f_{\rm adm} =0.1$, we assume that 10\% of the total dark matter is dark baryons. Comparing to the base model, 10\% of the base model CDM is replaced by dark baryons in the ADM model. In the following discussion, we will change $T_{{\rm d}\gamma}^0$ and $B_{\rm d}/T_{{\rm d}\gamma}^0$ separately to see the impacts on observables.

We also remind the reader that the matter fluctuation amplitude $\sigma_8$ is 
the rms fractional mass variance, computed in linear perturbation theory, of a mass field convolved with a tophat sphere of radius $8$ Mpc/$h$, where $h = H_0/(100 \; \textrm{km} \; \textrm{s}^{-1} \; \textrm{Mpc}^{-1})$. This is related to the matter power spectrum via
\begin{equation}
    \sigma_8^2 = \frac{1}{2\pi^2} \int k^3 P(k) |\widetilde{W}(k)|^2 \frac{dk}{k},
    \label{eq:sigma8}
\end{equation}
where $\widetilde{W}(k)$ is the Fourier transform of the tophat window function of radius 8 Mpc/$h$ and $P(k)$ is the linear-theory matter power spectrum today \cite[e.g.][]{Kolb:1990vq}.

\subsection{Altering the dark baryon-to-photon ratio} \label{sec:changeTd0}

\begin{figure*}
\centering
\includegraphics[trim={6cm 0.1cm 7cm 2.5cm},clip,width=\textwidth]{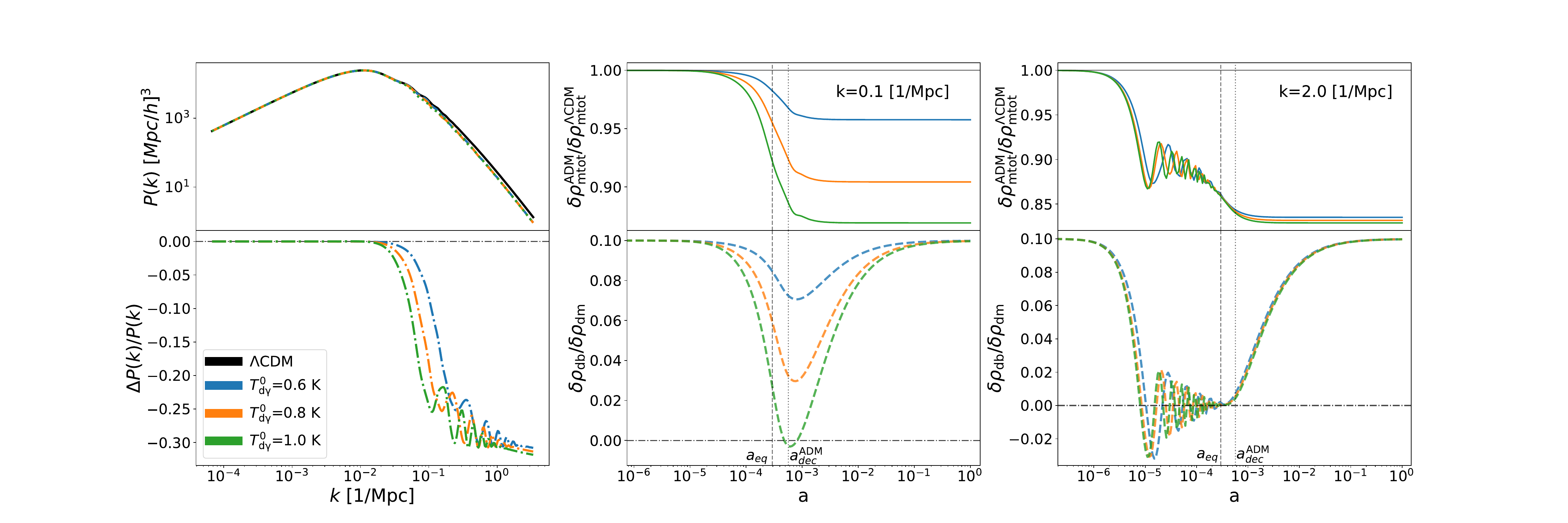}
\caption{Linear matter power spectra for the $\Lambda$CDM model and the ADM models with different dark photon temperatures today, $T_{{\rm d}\gamma}^0$, as well as the evolution of perturbations with $k=0.1 \ \mathrm{Mpc^{-1}}$ and $k=2 \ \mathrm{Mpc^{-1}}$. In the ADM models, we set $T_{{\rm d}\gamma}^0$ to 0.6 K, 0.8 K, and 1.0 K and keep the other parameters the same as in the fiducial ADM model in Eqs.~\eqref{eq:fidadm1} and \eqref{eq:fidadm2}. On the top of the middle and right panels, we show the ratios of the dimensional total matter perturbations between the ADM models and the base model. We show the ratio of dark baryon overdensity ($\delta \rho$) to the total dark matter overdensity of each model in the bottom part of the middle and the right panels. The vertical dashed (dotted) lines show the scale factor of matter-radiation equality (ADM decoupling).}
\label{fig:pwrspectraTdg_pk}
\end{figure*}

First, we investigate the impacts of changing $T_{{\rm d}\gamma}^0$, with all other parameters in Eqs.~\eqref{eq:basemodelpars}, \eqref{eq:fidadm1}, and \eqref{eq:fidadm2} held fixed at their fiducial values. Note that this means keeping $B_{\rm d}/T_{{\rm d}\gamma}^0$ fixed and, therefore, adjusting the binding energy accordingly.
By varying $T_{{\rm d}\gamma}^0$, we also effectively change $\eta_{\rm d}$, since the dark baryon density is fixed. 
To keep $N_{\rm eff}$ fixed while varying the dark photon temperature today, we adjust $N_\nu$ according to
\begin{equation}
    N_{\textrm{eff}} = N_\nu + \left(\frac{T_{{\rm d}\gamma}^0}{T_{\textrm{CMB}}^0}\right)^4 \frac{8}{7} \left(\frac{11}{4}\right)^\frac{4}{3},
    \label{eq:Neff}
\end{equation} 
where $T_{\textrm{CMB}}^0 \approx 2.7255$ K is the temperature of the CMB photons today \cite{fixsen_temperature_2009}. We look into three specific cases with dark photon temperature of 0.6 K, 0.8 K, and 1.0 K, which correspond to a dark photon energy density that is $0.34\% \ \mathrm{to} \ 2.62\%$ of the energy density in light relics.

\begin{figure*}
\centering
\includegraphics[trim={5.5cm 1cm 7cm 2.5cm},clip,width=\textwidth]{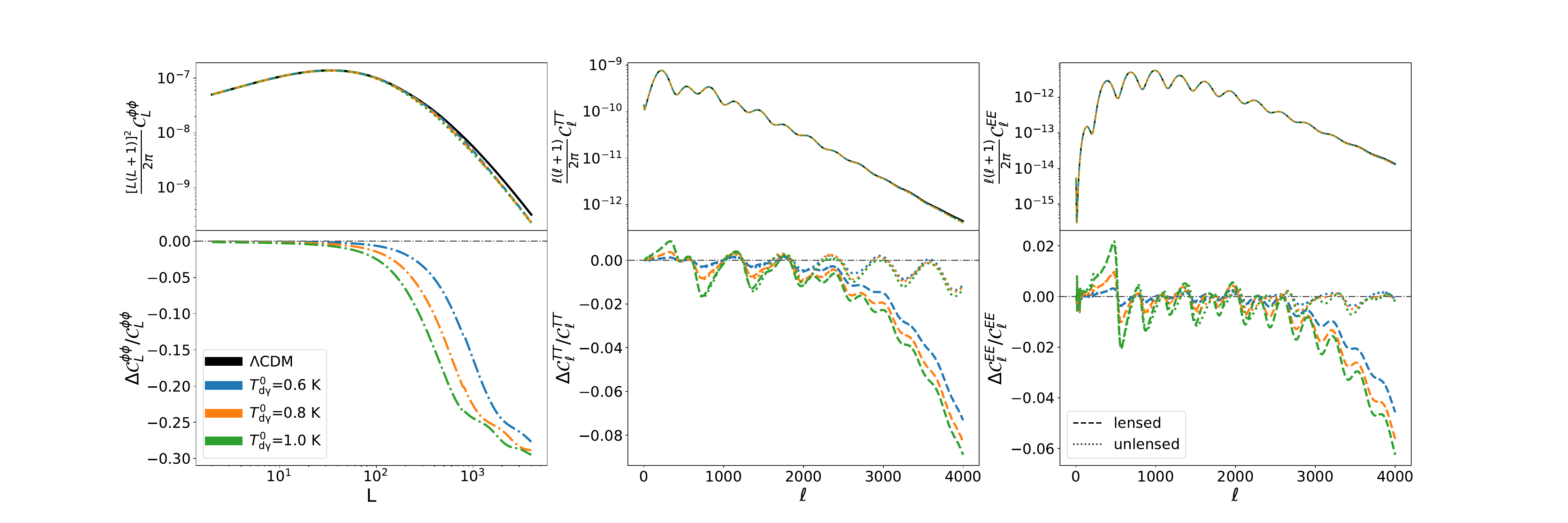}
\caption{Comparisons of CMB lensing, TT, and EE spectra between the $\Lambda$CDM model and the ADM models with three different values of $T_{{\rm d}\gamma}^0$. The ADM models and the fiducial \lcdm\ model are the same as those in Fig.~\ref{fig:pwrspectraTdg_pk}. We show from left to right in the top panels the CMB lensing potential, TT, and EE spectra. The fractional differences of these spectra with respect to the base \lcdm\ model are shown in the bottom panels. In the TT and EE fractional residual plots, dotted (dashed) lines are for unlensed (lensed) spectra.}
\label{fig:pwrspectraTdg_cmb}
\end{figure*}

As can be seen in Fig.~\ref{fig:pwrspectraTdg_cmb}, the most dramatic effects, on the TT and EE power spectra, of changing $T_{{\rm d}\gamma}^0$ are at $\ell \gtrsim 2000$ and are due to gravitational lensing. We thus consider these effects first, before moving on to studying the more subtle changes at $\ell \lesssim 2000$.

To understand the impact of gravitational lensing, we begin with the impact of varying $T_{{\rm d}\gamma}^0$ on the matter power spectrum, as shown in the left panel of Fig.~\ref{fig:pwrspectraTdg_pk}. At small scales, there is a suppression of power in ADM predictions compared to the base model prediction. This suppression is fundamentally due to the pressure support that the ADM receives from the dark photons, which slows the growth of $\delta \rho_{\rm adm}$. This in turn inhibits growth of $\delta \rho_{\rm cdm}$ as well, since the ADM is still contributing to the expansion rate (which acts against growth), while its contribution to gravitational potential gradients is suppressed. In the base model, all of the dark matter freely falls into the gravitational potential wells after horizon entry. However, in the ADM models, the freely falling CDM is just 90\% of the total dark matter (by mass), while 10\% of the dark matter mass is dark baryons, which receive at least some amount of pressure support until dark recombination. 

\begin{figure}[!]
    \includegraphics[width=\columnwidth]{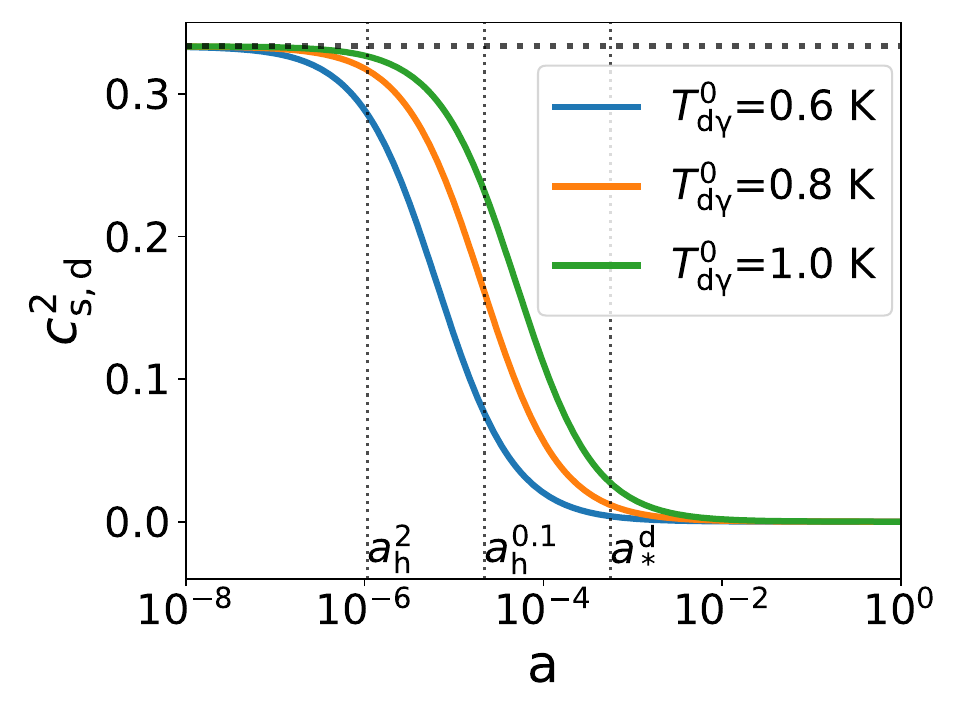}
    \caption{The dark sound speed $c_{\rm s,d}^2$ as a function of the scale factor for the indicated dark photon temperatures, with all other parameters set to those of our fiducial model. The horizon entries of modes with $k=2 \ \mathrm{Mpc^{-1}}$ and $k=0.1 \ \mathrm{Mpc^{-1}}$ are marked with $a_{\rm h}^2$ and $a_{\rm h}^{0.1}$, respectively. The scale factor at dark recombination is indicated by $a_*^{\rm d}$. At $a_*^{\rm d}$, the $k=3.25\times10^{-2} \ \mathrm{Mpc^{-1}}$ mode enters the horizon. Note that for this plot we ignore the impact of dark recombination, which guarantees the end of tight coupling and leads to a dramatic reduction of the pressure support felt by the dark baryons.}
    \label{fig:td_cs}
\end{figure}

The dependence of $P(k)$ suppression on $T_{{\rm d}\gamma}^0$ arises from the dependence of that pressure support on temperature. In the tight-coupling limit, pressure gradients are proportional to the square of the dark sound speed, which is given by
\begin{equation}
    c_{\rm s,d}^2 = \frac{1}{3}{\left(1+\frac{3\rho_{\rm db}}{4\rho_{{\rm d}\gamma}}\right)}^{-1},
\end{equation}
where 
\begin{equation}
    \frac{\rho_{\rm db}}{\rho_{{\rm d}\gamma}} \simeq 6.5 \left[\frac{10^4}{1+z}\right]\left[\frac{f_{\rm adm}}{0.1}\right]\left[\frac{0.8 \ \mathrm{K}}{T_{{\rm d}\gamma}^0}\right]^4.
\end{equation}
The evolution of $c_{\rm s,d}^2$ is shown in Fig.~\ref{fig:td_cs}. The dark sound speed is higher when the dark photon temperature is higher, which leads to greater pressure support of the ADM while it is tightly coupled to the dark photons. 

At large scales, the ADM predicted matter power spectra are similar to those predicted by the base model. This is because these modes enter the horizon after dark recombination, when dark baryons behave just like CDM because they have decoupled from the dark photons. The transition between these two regimes of large scales, where the ADM has no impact, and small scales, where there is suppression, is set by the wavenumber of the modes crossing the horizon during dark recombination, $k=3.25\times10^{-2} \ \mathrm{Mpc^{-1}}$.

To understand the impact on $P(k)$ in more detail, we show in the center and right panels of Fig.~\ref{fig:pwrspectraTdg_pk} the evolution of a mode in the transition region, with $k=0.1 \ \mathrm{Mpc^{-1}}$, and a mode more fully in the high-$k$ suppression region, with $k=2 \ \mathrm{Mpc^{-1}}$. In the top plots, we show the ratios of the dimensional perturbations of the total matter between the ADM models and the base model. For both modes, the dimensional perturbations of the total matter in the ADM models ($\delta \rho_{\rm mtot}^{\rm ADM}$) are suppressed compared to the base model prediction ($\delta \rho_{\rm mtot}^{\Lambda \rm CDM}$), and the ADM models with higher $T_{{\rm d}\gamma}^0$ show a larger suppression. For the different modes within the same ADM model, the deviation of $\delta \rho_{\rm mtot}^{\rm ADM}$ relative to $\delta \rho_{\rm mtot}^{\Lambda \rm CDM}$ at $a=1$ is larger for the $k=2 \ \mathrm{Mpc^{-1}}$ mode than for the $k=0.1 \ \mathrm{Mpc^{-1}}$ mode. This is caused by the different scale factor of horizon entry for the two modes. The $k=2 \ \mathrm{Mpc^{-1}}$ mode enters the horizon earlier than the $k=0.1 \ \mathrm{Mpc^{-1}}$ mode. Therefore, there is a longer duration of pressure support for the $k=2 \ \mathrm{Mpc^{-1}}$ mode. As a result, the ratio $\delta \rho_{\rm mtot}^{\rm ADM}/\delta \rho_{\rm mtot}^{\Lambda \rm CDM}$ is smaller at $a=1$ in the $k=2 \ \mathrm{Mpc^{-1}}$ mode than in the $k=0.1 \ \mathrm{Mpc^{-1}}$ mode of the same ADM model. Given the same background matter density in each model, this is consistent with the matter power spectra shown in the left panel of the figure.

For the $k=2 \ \mathrm{Mpc^{-1}}$ modes, horizon entry happens early when the sound speed of the dark plasma is still large. The dark baryons oscillate due to the dark photon pressure support, while the CDM perturbation is able to grow. As a result, the ratio $\delta \rho_{\rm db} / \delta \rho_{\rm dm}$ quickly decreases and oscillates around zero with decreasing amplitude. This is shown in the bottom right panel of Fig.~\ref{fig:pwrspectraTdg_pk}. At horizon entry, the sound speeds of the three ADM models are similarly large. As a result, the dark photon pressure-induced suppression on the gravitational potential is similar for all the three ADM models. The relative differences in $\Delta P(k)/P(k)$ and $\delta \rho_{\rm mtot}^{\rm ADM} / \delta \rho_{\rm mtot}^{\Lambda \rm CDM}$ among the three ADM models are small. 

For the $k=0.1 \ \mathrm{Mpc^{-1}}$ modes, horizon entry is later, occurring after a greater decrease in the dark sound speed than for the $k=2 \ \mathrm{Mpc^{-1}}$ modes. For these modes in the transition region, the decay is not only larger than in the $k=2 \ \mathrm{Mpc^{-1}}$ case, but the amount of decay depends significantly on $T_{{\rm d}\gamma}^0$. As shown in Fig.~\ref{fig:td_cs}, $c_{\rm s,d}^2$ for the $T_{{\rm d}\gamma}^0 = 0.6$ K model is below 0.1, while it is still above 0.2 for the $T_{{\rm d}\gamma}^0 = 1.0$ K model. The difference in dark pressure support is more significant at horizon entry among the three models. Additionally, the late horizon entry leads to a shorter duration of the dark acoustic oscillations, where dark recombination occurs before the completion of the first dark oscillation period. As a result, the fraction $\delta \rho_{\rm db} / \delta \rho_{\rm dm}$, as well as the amount of suppression in $\delta \rho_{\rm mtot}^{\rm ADM}$ compared to the $\delta \rho_{\rm mtot}^{\Lambda \rm CDM}$, is significantly different among the three models. This is shown in the middle panel in Fig.~\ref{fig:pwrspectraTdg_pk}. The $T_{{\rm d}\gamma}^0 = 0.6$ K model has the least amount of dark photon pressure. The dark baryons can cluster more easily with less pressure resistance. Therefore, the ADM model with $T_{{\rm d}\gamma}^0 = 0.6$ K has the largest $\delta \rho_{\rm adm}/\delta \rho_{\rm dm}$, as well as the smallest suppression in $\delta \rho_{\rm mtot}^{\rm ADM}/\delta \rho_{\rm mtot}^{\Lambda \rm CDM}$. As a result, the dispersion of $\Delta P(k) / P(k)$ is larger among the three ADM models for  modes with this wavenumber.

After understanding the differences in the matter power spectra, we now turn to the CMB lensing spectra of the ADM models. In the left panel of Fig.~\ref{fig:pwrspectraTdg_cmb}, we see that at small scales the lensing potential spectrum amplitudes predicted by the ADM models are lower than the base model amplitude predictions, with a higher $T_{\rm d\gamma}^0$ leading to a larger suppression. This is consistent with the changes seen in the matter power spectra. Recall that the lensing potential spectrum \cite{Lewis:2006fu} is 
\begin{equation}
    C_L^{\phi\phi} \propto \int \frac{dk}{k} \left[\int_0^{\chi_*} d\chi \Phi(k; \eta_0 - \chi) j_L(k\chi) \left(\frac{\chi_* - \chi}{\chi_* \chi}\right) \right]^2,
\end{equation}
where $\chi$ is the comoving distance, $\eta_0$ is the conformal time today, $j_L$ is the spherical Bessel function, $\Phi$ is the Weyl potential, and the subscript $*$ indicates recombination. Most of the lensing effects come from redshift $z \lesssim 20$ \cite[e.g.][]{Lewis:2006fu}, deep in the matter-dominated era. Therefore, we can use the Poisson equation to replace $\Phi$ with $\delta_{\rm mtot}$ via
\begin{equation}
    \Phi(k, z) = -\frac{3}{2}\Omega_{\rm m} H_0^2 (1+z) k^{-2} \delta_{\rm mtot}(k,z).
\end{equation}
Given that the background cosmology is the same in the ADM models and in the base model, the differences in the CMB lensing potential spectra are caused by the matter power spectra differences. Comparing the matter power spectra and the lensing potential spectra, this expected correlation is seen in Figs.~\ref{fig:pwrspectraTdg_cmb} and \ref{fig:pwrspectraTdg_pk}.

The CMB TT and EE spectra are shown in the middle and right panels of Fig.~\ref{fig:pwrspectraTdg_cmb}. In both panels, the fractional changes relative to the base model predictions for the lensed and unlensed spectra are shown. We find that the main effects on the lensed TT and EE spectra are caused by CMB lensing. The unlensed ADM spectra deviate only mildly from base model predictions, but the lensed spectra deviate significantly at small scales. This is expected due to the mixing of power at different scales by CMB lensing \cite{Lewis:2006fu}. At small scales, where the unlensed CMB spectra are close to zero due to diffusion damping, the lensing effect of power mixing provides the main contribution to the amplitude of the lensed spectra. The lensing potential amplitude is smaller when the dark photon temperature is higher, leading to lower mixing power. As a result, the lensed CMB spectra with lower $T_{{\rm d}\gamma}^0$ deviate less from the base model prediction. 

\begin{figure*}
\centering
\includegraphics[trim={6cm 0.1cm 7cm 2.5cm},clip,width=\textwidth]{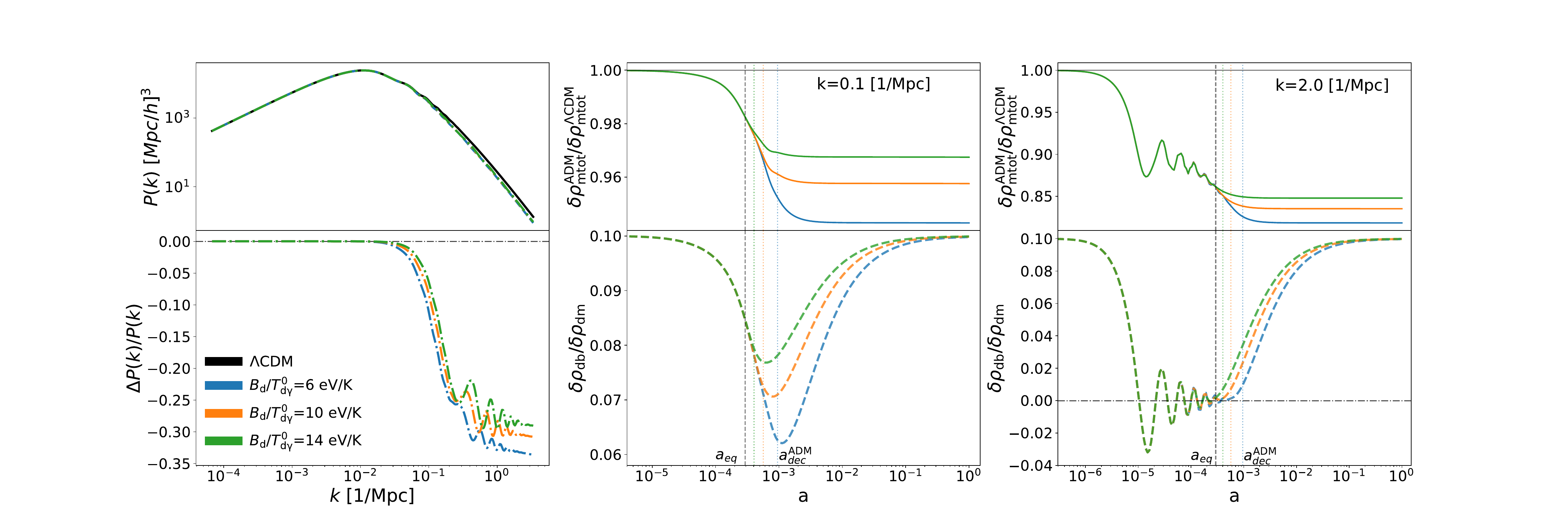}
\caption{Similar to Fig.~\ref{fig:pwrspectraTdg_pk}. The dark photon temperature is fixed in all the ADM models to $T_{{\rm d}\gamma}^0 = 0.6 \ {\rm K}$, and the ratio of dark hydrogen binding energy-to-dark photon temperature is set to either $B_{\rm d}/T_{{\rm d}\gamma}^0 = 6, \ 8, \ {\rm or} \ 10 \ \mathrm{eV/K}$. The dashed vertical lines show the scale factor of matter-radiation equality. The colored vertical dotted lines show the scale factor of ADM decoupling for the different ADM models.}
\label{fig:pwrspectraBd_pk}
\end{figure*}

\begin{figure*}
\centering
\includegraphics[trim={5.5cm 1cm 7cm 2.5cm},clip,width=\textwidth]{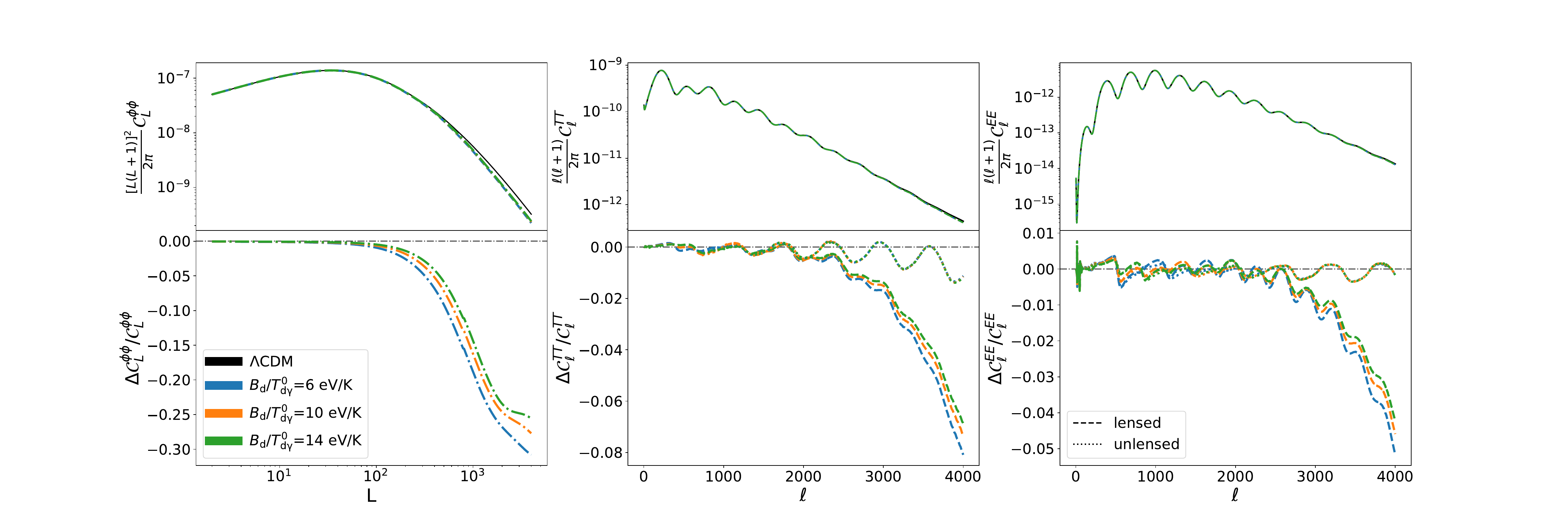}
\caption{Similar to Fig.~\ref{fig:pwrspectraTdg_cmb}. The dark photon temperature is fixed in all the ADM models to $T_{{\rm d}\gamma}^0 = 0.6 \ {\rm K}$, and the ratio of dark hydrogen binding energy-to-dark photon temperature is set to either $B_{\rm d}/T_{{\rm d}\gamma}^0 = 6, \ 8, \ {\rm or} \ 10 \ \mathrm{eV/K}$.}
\label{fig:pwrspectraBd_cmb}
\end{figure*}

We now turn to the differences at $\ell\lesssim2000$ where CMB lensing effects are less important. Although we have not developed a complete analytic understanding of the observed changes to the unlensed TT and EE spectra, we do know they are driven by changes that occurred before and during last scattering, and we have isolated the physical effects that must be at play. The changes to the spectra all follow from changes to two quantities:
i) the fraction of free-streaming radiation ($R_\nu = \rho_\nu / \rho_{\rm rad}$), and ii) the fraction of dark matter that is pressure supported. Increasing $T_{{\rm d}\gamma}^0$ reduces $R_\nu$, and increases the pressure-supported dark matter fraction. The former boosts the superhorizon gravitational potential and photon monopole amplitudes, while the former and the latter both influence subhorizon evolution of gravitational potentials, and therefore the acoustic oscillations in the plasma.  A shift in the force-free point of the acoustic oscillations and a temporal phase shift can produce spectral differences similar to the ones we see here. Effects of varying $R_\nu$ are described in Ref.~\cite{Bashinsky:2003tk}, while the impact of the pressure support change is described in Ref.~\cite{ge_scaling_2022}.

\subsection{Changing the dark binding energy-to-dark photon temperature ratio}\label{sec:changeBd}

\begin{centering}
\begin{table*}
\begin{ruledtabular}
\begin{tabular}{ccc}
    Label & Datasets & BBN consistency? \\ \hline
    BBN consistent, with SH0ES data & \emph{Planck}, BAO, SH0ES & yes\\
    BBN consistent, without SH0ES data & \emph{Planck}, BAO & yes\\
    BBN inconsistent, with SH0ES data & \emph{Planck}, BAO, SH0ES & no\\
    BBN inconsistent, without SH0ES data & \emph{Planck}, BAO & no\\
\end{tabular}
\end{ruledtabular}
\caption{Summary of datasets and constraints used in this paper. CMB data are from the \emph{Planck} 2018 high-$\ell$ TT+TE+EE, low-$\ell$ TT, low-$\ell$ EE, and lensing datasets \cite{planck_collaboration_planck_2020}; BAO data are from 6dFGS \cite{beutler_6df_2011}, SDSS MGS \cite{ross_clustering_2015}, and BOSS DR12 \cite{alam_clustering_2017}; and the SH0ES $H_0$ value is from Ref.~\cite{riess_comprehensive_2022}. BBN consistency indicates constraints on helium abundance from big bang nucleosynthesis (BBN).}
\label{tab:datasets}
\end{table*}
\end{centering}

In this subsection, we investigate the physical effects on the matter and CMB power spectra of altering the dark binding energy-to-dark photon temperature ratio ($B_{\rm d}/T_{{\rm d}\gamma}^0$). We adjust $B_{\rm d}/T_{{\rm d}\gamma}^0$ by varying $B_{\rm d}$ while keeping all other parameters in Eqs.~\eqref{eq:basemodelpars}, \eqref{eq:fidadm1}, and \eqref{eq:fidadm2} fixed. We show spectra for models with $B_{\rm d}$ set to 3.6, 6, and 8.4 eV. The pressure in the dark baryon-photon plasma prior to dark recombination is the same for all of these models since they all have the same dark photon temperature. However, varying $B_{\rm d}/T_{{\rm d}\gamma}^0$ changes the time of dark recombination and, therefore, changes the time at which this pressure support precipitously drops.

As in Sec.~\ref{sec:changeTd0}, we first discuss the matter power spectra. In the left panel of Fig.~\ref{fig:pwrspectraBd_pk}, we show the matter power spectrum predictions for the ADM models with different $B_{\rm d}/T_{{\rm d}\gamma}^0$, as well as the prediction from the base model. At large scales, at which the modes enter the horizon after dark recombination, the ADM predicted matter power spectra are identical to those predicted by the base model. At small scales, ADM predicted matter power spectra are suppressed relative to the base model prediction.

The key factor leading to the difference in $\Delta P(k) / P(k)$ is the duration of the pressure support between horizon entry and dark recombination. We pick two modes ($k=0.1\ \mathrm{Mpc^{-1}}$ and $k=2\ \mathrm{Mpc^{-1}}$ in the middle and right panels of Fig.~\ref{fig:pwrspectraBd_pk}, respectively) to see the perturbation evolution. When the dark binding energy is higher, more energy is required for a dark photon to photoionize a bound state, leading to an earlier dark recombination epoch. After dark recombination, dark photons begin to free stream, and the ADM pressure support drops to near zero, stopping any further suppression of the total matter perturbations compared to the base model prediction. For the same mode, the perturbation evolutions before dark recombination are quite similar among the three ADM models. This is expected because the dark photon pressure amplitudes are the same when the dark photon temperature is fixed. The differences only appear after dark recombination has occurred in one of the models. For the model with dark recombination at a later scale factor, the pressure support lasts longer, leading to a longer suppression of the clustering of ADM despite the same background evolution as other ADM models. Thus, $\delta \rho_{\rm mtot}$ is further suppressed, leading to a smaller $\Delta P(k) / P(k)$. 

As for dependence on scale, smaller scales enter the horizon earlier, leading to a longer duration of pressure support between horizon entry and dark recombination. Thus, the total matter perturbation is more suppressed and the matter power is lower as the scale becomes smaller. This scale dependence becomes quite weak for modes that enter the horizon deep in the radiation-dominated era since matter is a small fraction of the total density. 

In Fig.~\ref{fig:pwrspectraBd_cmb}, we show the CMB TT, EE, and lensing potential spectra with three different dark binding energies, $B_{\rm d} = 3.6 \ \mathrm{eV}, \ 6 \ \mathrm{eV} \ {\rm and} \ 8.4 \ \mathrm{eV}$, and compare them to the fiducial \lcdm\ model. For similar reasons to those discussed in Sec.~\ref{sec:changeTd0}, the CMB lensing potential spectra are closely related to the matter power spectra. Comparing the lensed and unlensed TT and EE spectra, we find that CMB lensing is the main driver of the differences in the lensed CMB spectra at small scales.

\section{Constraints placed by current data} \label{sec:constraints}

\begin{figure*}
\centering
     \begin{subfigure}[b]{0.49\textwidth}
         \centering
         \includegraphics[width=\textwidth]{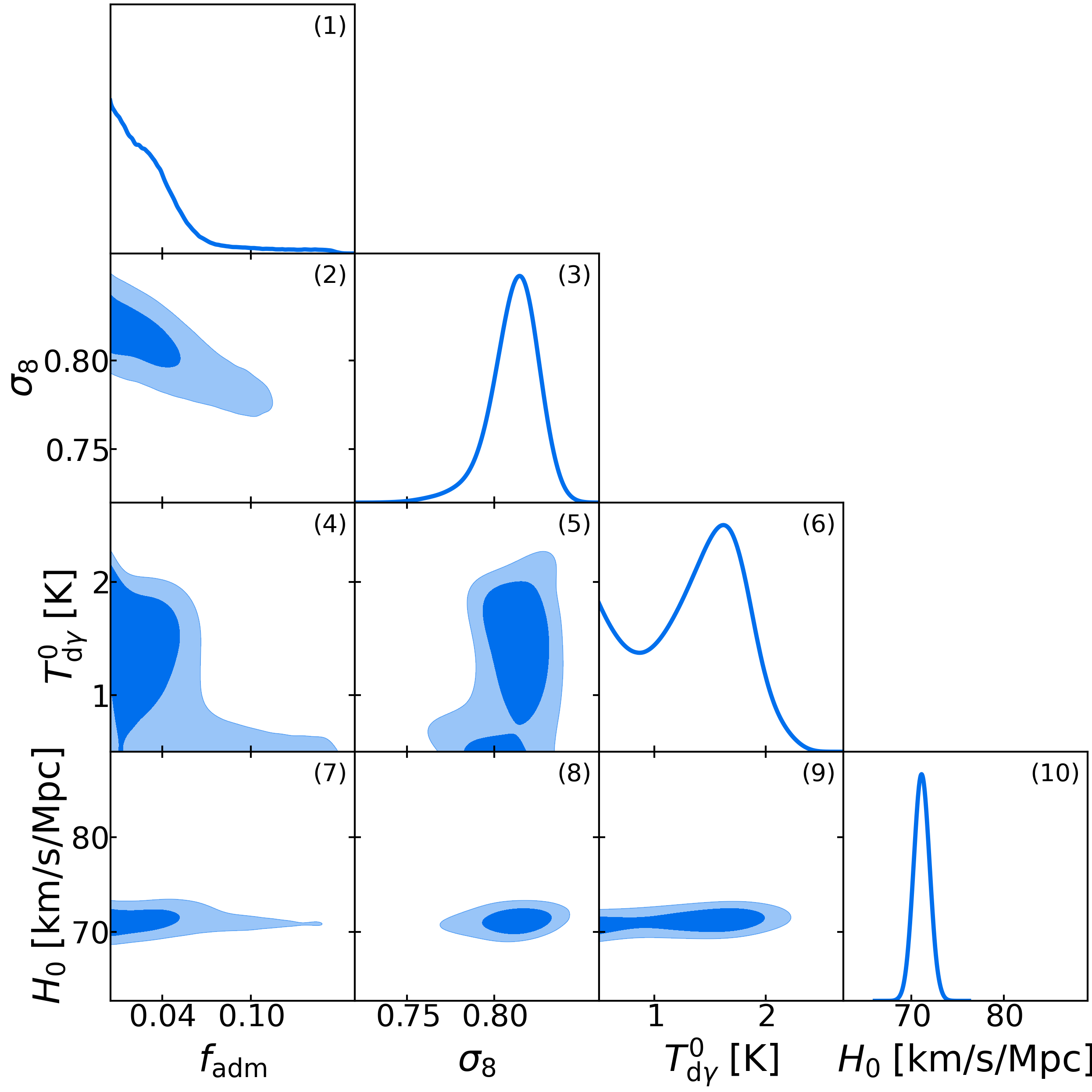}
         \caption{BBN consistent, with SH0ES data}
         \label{fig:mcmc1}
     \end{subfigure}
     \hfill
     \begin{subfigure}[b]{0.49\textwidth}
         \centering
         \includegraphics[width=\textwidth]{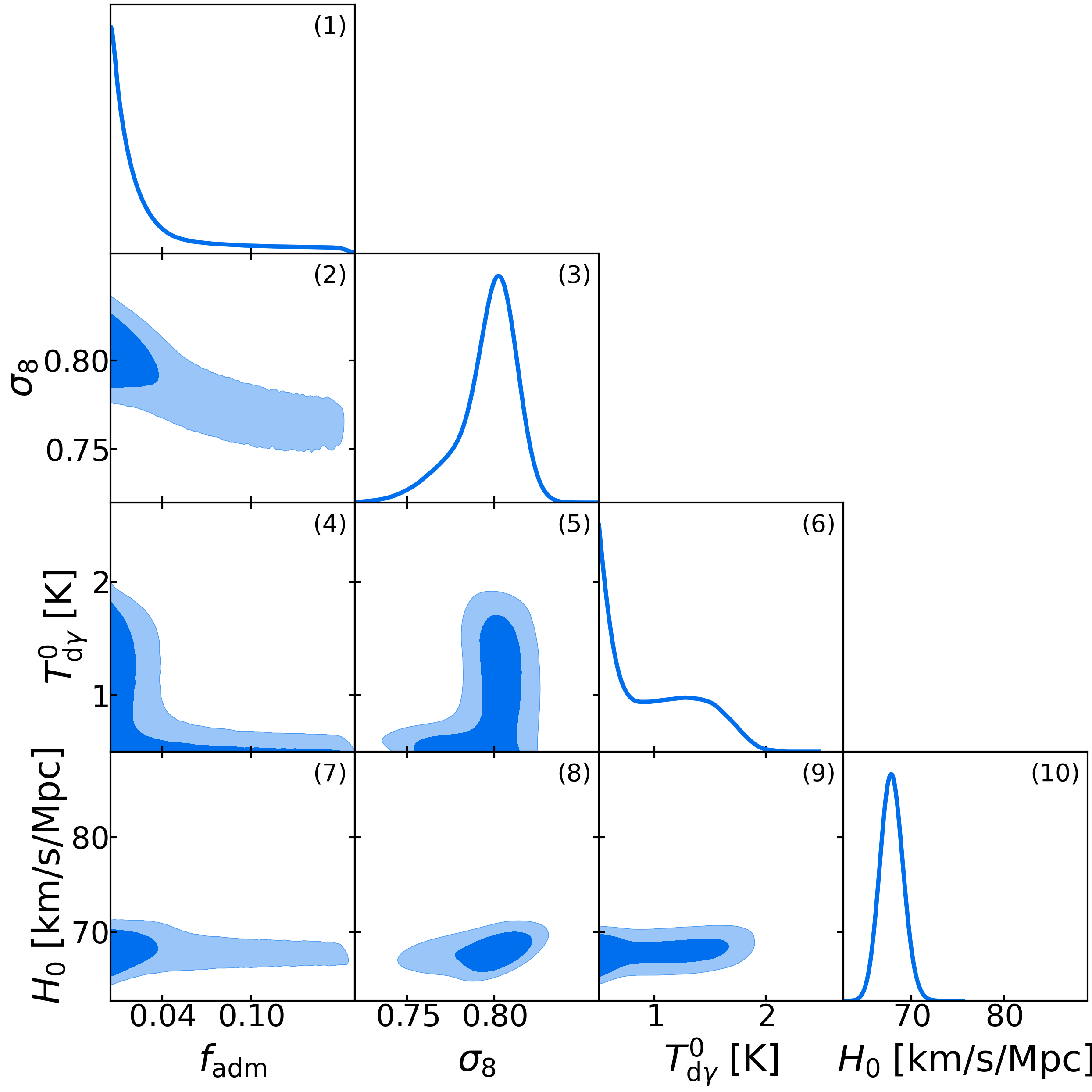}
         \caption{BBN consistent, without SH0ES data}
         \label{fig:mcmc2}
     \end{subfigure}
     \hfill
     \vfill
     \begin{subfigure}[b]{0.49\textwidth}
         \centering
         \includegraphics[width=\textwidth]{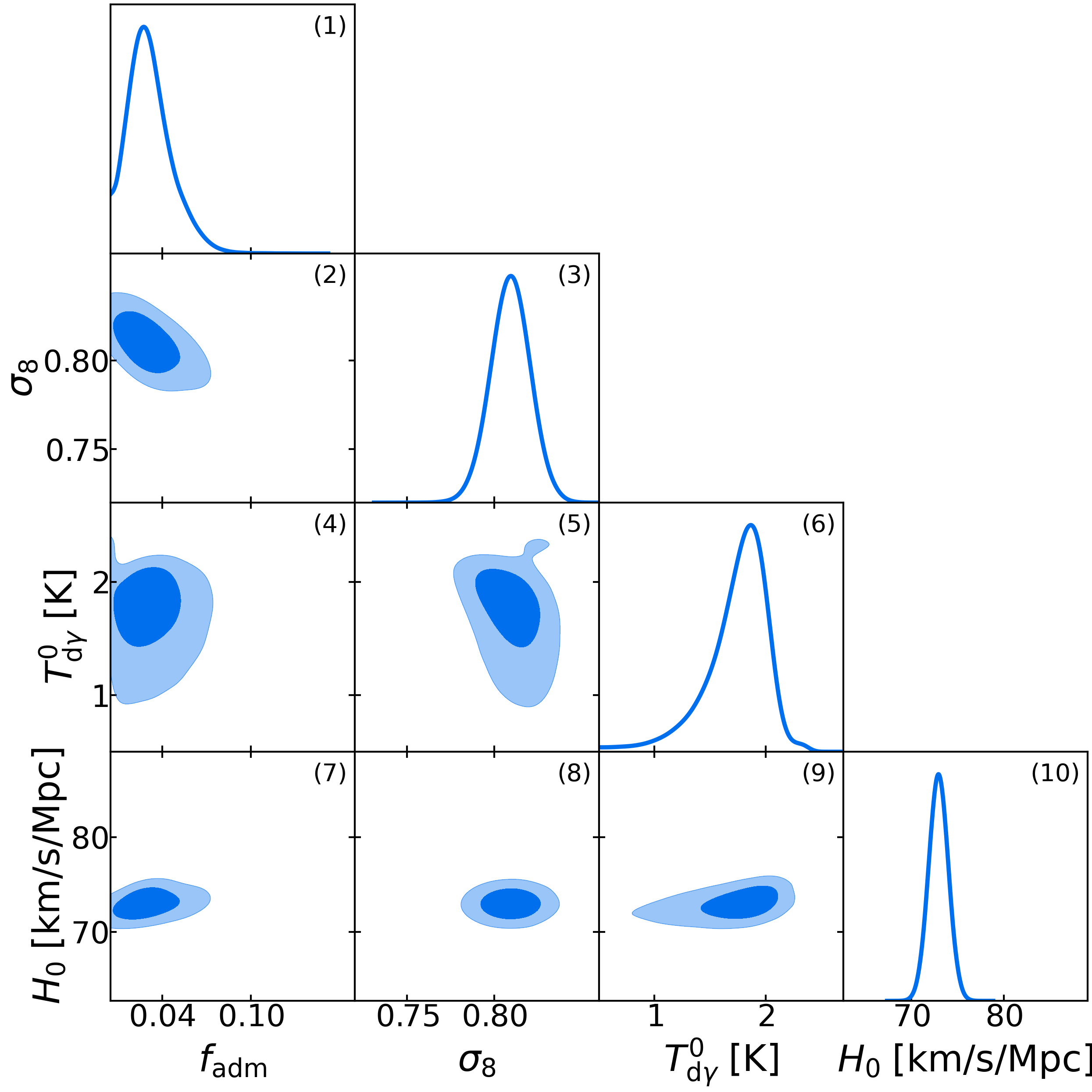}
         \caption{BBN inconsistent, with SH0ES data}
         \label{fig:mcmc3}
     \end{subfigure}
     \hfill
     \begin{subfigure}[b]{0.49\textwidth}
         \centering
         \includegraphics[width=\textwidth]{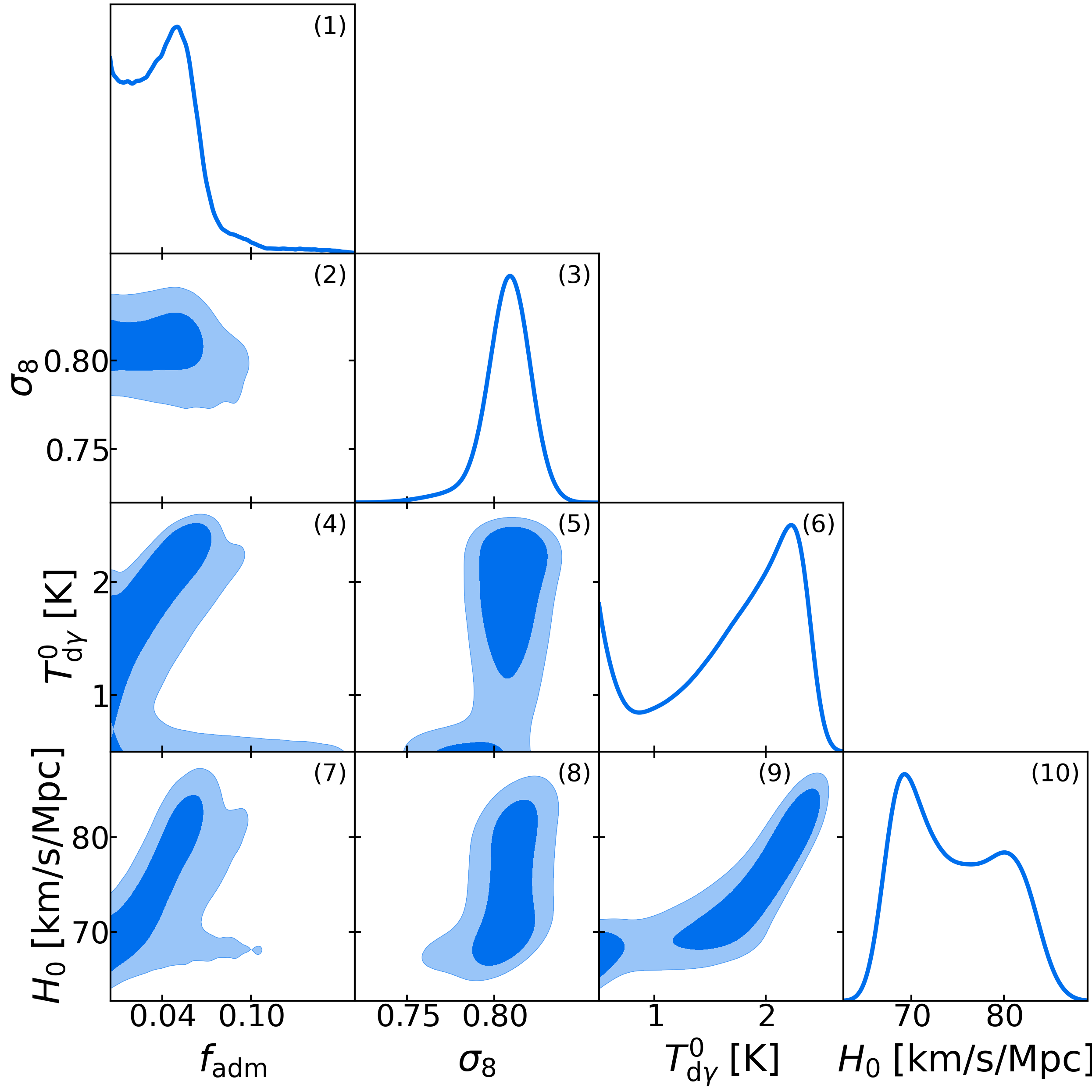}
         \caption{BBN inconsistent, without SH0ES data}
         \label{fig:mcmc4}
     \end{subfigure}
     \hfill
\caption{1- and 2-dimensional marginal posterior probability densities for $f_{\textrm{adm}}$, $\sigma_8$, $T_{{\rm d}\gamma}^0$, and $H_0$ for all four model spaces. The diagonal shows the (arbitrarily normalized) 1-dimensional posteriors. The inner and outer contours on the 2-dimensional posteriors enclose the 68\% and 95\% credible regions, respectively.}
\label{fig:2dposteriors}
\end{figure*}

In this section, we explore the parameter constraints placed by current observational data on the ADM model described in Sec.~\ref{sec:modelspace}. The full cosmological parameter set is 
\begin{equation}
     \{H_0, \Omega_{\rm b} h^2, \Omega_{\rm c} h^2, A_{\rm s}, n_{\rm s}, \tau, N_\nu, T_{{\rm d}\gamma}^0, \Omega_{\rm db}h^2, B_{\rm d}/T_{{\rm d}\gamma}^0\},
\end{equation}
which is a combination of \lcdm\ + $N_{\rm eff}$ plus three ADM parameters. We let $N_{\rm eff}$ vary freely in addition to the dark photon temperature $T_{{\rm d}\gamma}^0$ via Eq.~\eqref{eq:Neff}. In practice we implement this freedom by allowing the number of light neutrino species $N_\nu$ to adjust away from its \lcdm\ value of 3.044 \cite{Bennett:2020zkv,Akita_2020,Froustey:2020mcq}. We included this freedom so that our model space would contain the scaling transformation direction highlighted by Refs.~\cite{cyr-racine_symmetry_2022} and \cite{ge_scaling_2022}, and its associated large posterior $H_0$ uncertainty. In the explored models, we consider both cases where the primordial helium abundance is predicted by big bang nucleosynthesis (BBN) and where it is a free parameter.

We use CMB data from \emph{Planck}, namely the \emph{Planck} 2018 high-$\ell$ TT+TE+EE, low-$\ell$ TT, low-$\ell$ EE, and lensing datasets \cite{planck_collaboration_planck_2020}, as well as BAO data from 6dFGS \cite{beutler_6df_2011}, SDSS MGS \cite{ross_clustering_2015}, and BOSS DR12 \cite{alam_clustering_2017}. Additionally, we consider parameter constraints both with and without the inclusion of the independent 2022 measurement of $H_0$ from SH0ES \cite{riess_comprehensive_2022}. While Ref.~\cite{riess_comprehensive_2022} is no longer the latest SH0ES result, it was the most recent data available at the time we began this work. Hereafter, SH0ES data will refer to Ref.~\cite{riess_comprehensive_2022}. The datasets and constraints used in the following analyses to obtain joint constraints on parameter values are summarized in Table \ref{tab:datasets}.

We use a modified version of CAMB \cite{lewis_efficient_2000, cyr-racine_cosmology_2013, Cyr-Racine:2013fsa}, a publicly available Einstein-Boltzmann solver, to make model predictions and CosmoMC \cite{lewis_cosmological_2002} to calculate the posterior probability distributions of the parameters using Markov chain Monte Carlo (MCMC) methods.
We applied a flat prior on the ADM parameters as
\begin{gather}
    T_{{\rm d}\gamma}^0 \ [{\rm K}] \in [0.5, 2.7], \ \Omega_{{\rm db}}h^2 \in [0.0001, 0.02], \notag \\ {\rm and} \ B_{\rm d} / T_{{\rm d}\gamma}^0 \ [{\rm eV/K}] \in [2.5, 15].
\end{gather}
This choice of prior on $B_{\rm d} / T_{{\rm d}\gamma}^0$ allows dark recombination to occur at a wide range of redshifts approximately between 475 and 2700. We set the energy densities of the dark photons and dark baryons to not exceed those of the photons and baryons in the visible sector defined in the base model in Sec.~\ref{sec:modelspace}. The lower bound on $T_{{\rm d} \gamma}^0$ is due to a computational limitation, where a lower $T_{{\rm d}\gamma}^0$ occasionally results in a failure of the ODE solver for dark ionization history. However, at the lowest allowed $T_{{\rm d}\gamma}^0$, pressure support is significantly reduced (relative to the maximal high-temperature amount) at sufficiently early times to allow for the emergence of the low $\sigma_8$ solution that is a major focus of this paper. We leave improvements to address this numerical difficulty for future work.

In Sec.~\ref{sec:scalingsym}, we discuss the emergence, in our constraint contours for the BBN-inconsistent model without SH0ES data, of the scaling transformation of Ref.~\cite{cyr-racine_symmetry_2022}. In Sec.~\ref{sec:sigma8andTdg0}, we examine the ADM parameter space at low $T_{{\rm d}\gamma}^0$ and find that lower values of $\sigma_8$ are allowed in this regime. In Sec.~\ref{sec:H0}, we explore the possibility of easing the $H_0$ tension with ADM.

\subsection{Emergence of the scaling symmetry solution} \label{sec:scalingsym}

The authors of Ref.~\cite{cyr-racine_symmetry_2022} introduced a scaling of rates in the Einstein-Boltzmann equations, and associated scaling of the amplitude of the primordial power spectrum, that leaves dimensionless cosmological observables invariant. This FFAT (for Free-Fall rate, Amplitude, and Thomson rate) scaling transformation is given by 
\begin{flalign}
    \sqrt{G \rho_i(a)} &\rightarrow \lambda \sqrt{G \rho_i(a)}, \notag \\
    \sigma_{\textrm{T}} n_e(a) &\rightarrow \lambda \sigma_{\textrm{T}} n_e(a), \ {\rm and} \notag \\
    A_{\rm s} &\rightarrow A_{\rm s}/\lambda^{(n_{\rm s}-1)},
    \label{eq:scalings}
\end{flalign}
where $a = 1/(1+z)$ is the scale factor and $\lambda$ is the scaling transformation parameter \cite{cyr-racine_symmetry_2022}. 

As explained in Ref.~\cite{cyr-racine_symmetry_2022}, the FFAT scaling transformation can be mimicked in a model with ADM, additional free-streaming light relics, and free helium. Thus, when we allow for BBN inconsistency (so that helium is free) and do not restrict the value of $H_0$ with use of SH0ES (so as not to pin down $\lambda$ too tightly), parameters are free to move along a subspace of the expanded parameter space we use here, corresponding to this FFAT-mimicking scaling transformation, with little to no impact on the dimensionless observables in our likelihood. 

We therefore expect to see this parameter-space direction emerge in the constant probability density contours of our BBN-inconsistent no-SH0ES posterior shown in Fig.~\ref{fig:mcmc4}. The elongated features in Figs.~\ref{fig:mcmc4}(4), \ref{fig:mcmc4}(7), and \ref{fig:mcmc4}(9) indicate a positive correlation between $f_{\textrm{adm}}$ and $T_{{\rm d}\gamma}^0$, between $f_{\textrm{adm}}$ and $H_0$, and between $T_{{\rm d}\gamma}^0$ and $H_0$, respectively, consistent with the expectations due to the scaling transformation symmetry. 

As pointed out in Ref.~\cite{cyr-racine_symmetry_2022}, the scaling transformation, at least as they implement it, requires BBN-inconsistent helium abundances. Also noteworthy, the helium abundance required for consistency with SH0ES is significantly below determinations from observations \cite{aver2021, Izotov:2014fga,Aver:2015iza,2018MNRAS.478.5301F,Hsyu:2020uqb}. These are formidable challenges to the building of a model that can successfully exploit the FFAT scaling to solve the Hubble tension (see, however, Ref.~\cite{Greene:2023cro}). Nevertheless, we see the explanatory power of the FFAT symmetry, as it explains these features of the posterior.

\subsection{Constraints at low {\boldmath$T_{{\rm d}\gamma}^0$}} \label{sec:sigma8andTdg0}

As expected from Sec.~\ref{sec:changeTd0}, the deviations from \lcdm\ in the CMB power spectra increase with higher $T_{{\rm d}\gamma}^0$ across the angular scales measured well by \emph{Planck} ($\ell\lesssim2000$). With low $T_{{\rm d}\gamma}^0$, dark photons provide only very mild pressure support to dark baryons (and almost none at all after dark recombination), leading to small deviations in the CMB power spectra. Thus, at low $T_{{\rm d}\gamma}^0$, ADM is semi-degenerate with CDM, and the ADM parameter space opens up significantly as a result, as can be seen with the higher values of $f_{\rm adm}$ allowed in Figs.~\ref{fig:mcmc1}(4), \ref{fig:mcmc2}(4), and \ref{fig:mcmc4}(4).

However, there is not a complete degeneracy between ADM and CDM at low $T_{{\rm d}\gamma}^0$, and distinct observational signatures between the two models remain. As discussed in Sec.~\ref{sec:changeTd0}, even with low $T_{{\rm d}\gamma}^0$, the matter power spectrum in the presence of ADM is damped relative to the $\Lambda$CDM matter power spectrum. Therefore, lower values of $\sigma_8$ in line with the KiDS-1000 result \cite{heymans_kids-1000_2021} are allowed with lower $T_{{\rm d}\gamma}^0$, as can be seen in Figs.~\ref{fig:mcmc1}(5), \ref{fig:mcmc2}(5), and \ref{fig:mcmc4}(5).

Furthermore, a higher fraction of pressure-supported dark matter also yields a dampened matter power spectrum, and therefore a reduced $\sigma_8$, relative to that of \lcdm. Thus, the above-mentioned tolerance for higher $f_{\rm adm}$ at low $T_{{\rm d}\gamma}^0$ also allows for lower $\sigma_8$, as can be seen in Figs.~\ref{fig:mcmc1}(2), \ref{fig:mcmc2}(2), and \ref{fig:mcmc3}(2). This low $\sigma_8$ and low $T_{{\rm d}\gamma}^0$ region of parameter space is interesting because it could provide a solution to the $\sigma_8$ tension. We discuss this potential solution more below in Sec.~\ref{sec:lowsigma8}. As we will see in Sec.~\ref{sec:forecastresults}, this mechanism for reducing $\sigma_8$ appears to be testable with future CMB measurements.

We now compare with another recent exploration of the ADM model parameter space \cite{bansal_precision_2022}. In addition to $f_{\rm adm}$, $T_{{\rm d}\gamma}^0$, and $B_{\rm d}/T_{{\rm d}\gamma}^0$, they also freed up the dark proton mass and dark fine structure constant. Somewhat surprisingly, since our model space is a subspace of theirs, their marginal posteriors, when using the datasets we use, do not indicate the presence of a low $\sigma_8$ solution. It is possible this is due to differential volume effects as they have two additional dimensions over which to marginalize. For a recent discussion of volume effects and how they can cause interesting solutions to get hidden (if one is ``looking" with the posterior probability distribution), see Ref.~\cite{meiers2023}. Some support for this hypothesis comes from the fact that when they adopt a strong prior on a low $\sigma_8$ via the inclusion of KiDS-1000 data \cite{heymans_kids-1000_2021}, they find their model can accommodate it. 

\begin{figure*}
    \centering
    \includegraphics[width=\textwidth]{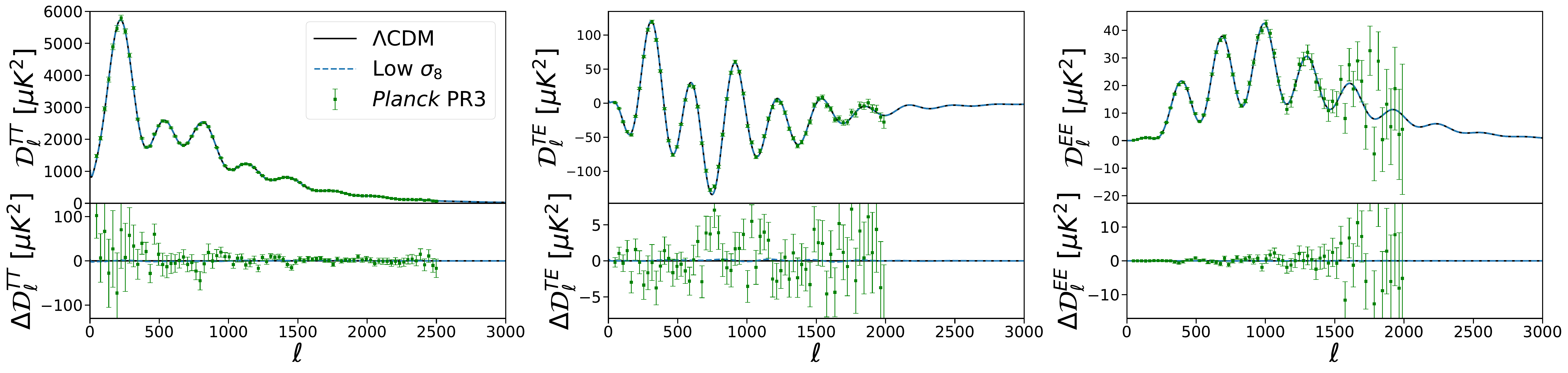}
    \includegraphics[width=0.67\textwidth]{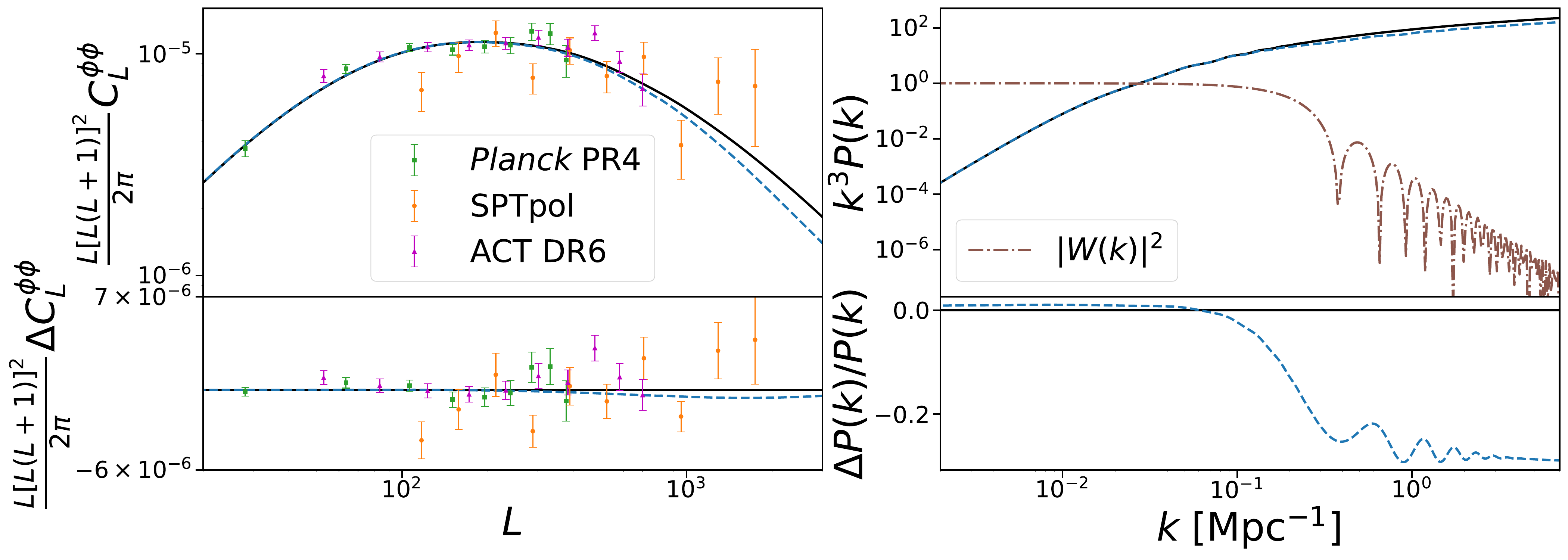}
    \caption{Primary CMB, CMB lensing, and linear matter power spectra for the best-fit BBN-consistent model with $T_{{\rm d}\gamma}^0 < 0.75$, $f_{\textrm{adm}} > 0.09$, and fixed $N_\nu = 3.044$. In this model, plotted in the blue dashed lines, 
    $H_0 = 67.52$ $\textrm{km} \; \textrm{s}^{-1} \; \textrm{Mpc}^{-1}$, $\Omega_{\rm b} h^2 = 0.02241$, $\Omega_{\rm c} h^2 = 0.1076$, $\tau = 0.05768$, $A_{\rm s} = 2.113 \times 10^{-9}$, $n_{\rm s} = 0.9667$, $f_{\textrm{adm}} = 0.1024$, $B_{\rm d} = 7.478$ eV, $T_{{\rm d}\gamma}^0 = 0.5001$ K, $N_\nu = 3.044$, and $\sigma_8 = 0.7869$. The solid black lines show the fiducial $\Lambda$CDM power spectra generated using the best fits from Ref.~\cite{planck_collaboration_planck_2020-1}. Top row: TT, TE, and EE spectra from the left to right, where $\mathcal{D}_{\ell} = \ell (\ell + 1) C_\ell (T_{\rm CMB}^0)^2 / (2 \pi)$. The green data points on the TT, TE, and EE power spectra show the \emph{Planck} PR3 2018 data \cite{planck_collaboration_planck_2020}. Bottom left: Lensing power spectra with \emph{Planck} PR4 data \cite{planck_collaboration_planck_2020_PR4} shown in green, SPTpol data \cite{bianchini_constraints_2020} shown in orange, and ACT DR6 data \cite{qu_atacama_2023, madhavacheril_atacama_2023} shown in magenta. Bottom right: Linear matter power spectra with $|W(k)|^2$ appropriate for $\sigma_8$ shown in brown; see Eq.~\eqref{eq:sigma8}.}
    \label{fig:lowsigma8pwrspectra}
\end{figure*}

\subsection{Constraints on {\boldmath$H_0$}} \label{sec:H0}

As discussed above in Sec.~\ref{sec:scalingsym}, for the BBN-inconsistent model without the SH0ES data in Fig.~\ref{fig:mcmc4}, a broad range of $H_0$ values is allowed by current CMB and BAO data as a result of the FFAT scaling transformation symmetry in Ref.~\cite{cyr-racine_symmetry_2022}. Of course, for the cases where we include SH0ES data, the $H_0$ uncertainty decreases dramatically. Even without including SH0ES data, enforcing BBN consistency prevents the Thomson rate from following FFAT scaling, also leading to a dramatic reduction of the $H_0$ posterior width.

While higher values of $H_0$ are not preferred by current CMB and BAO data for the BBN-consistent model without the SH0ES data in Fig.~\ref{fig:mcmc2}(10), tolerance for higher values of $H_0$, in line with the latest SH0ES measurement \cite{murakami_leveraging_2023}, is well above what we have in the case of \lcdm. Thus, the partial opening of the ADM parameter space in this work still could provide a possible, albeit partial, resolution to the $H_0$ tension, as seen in Ref.~\cite{cyr-racine_symmetry_2022}.

It is worth noting that, for the BBN-consistent model without the SH0ES data, higher $H_0$ and lower $\sigma_8$ are not simultaneously allowed by current data, as seen in Fig.~\ref{fig:mcmc2}(8). Similarly, the addition of the SH0ES $H_0$ data in Fig.~\ref{fig:mcmc1}(3) tightens the 1-dimensional constraints on $\sigma_8$, as compared to the constraints without the SH0ES data in Fig.~\ref{fig:mcmc2}(3), thereby excluding the lowest values of $\sigma_8$. This effect of partially excluding low $\sigma_8$ with the inclusion of the SH0ES data can also be seen in the 2-dimensional constraints on $\sigma_8$ and $T_{{\rm d}\gamma}^0$ in Fig.~\ref{fig:mcmc1}(5). Therefore, the low $\sigma_8$ and low $T_{{\rm d}\gamma}^0$ region of parameter space discussed above in Sec.~\ref{sec:sigma8andTdg0} does not provide a simultaneous solution to both the $H_0$ and $\sigma_8$ tensions.

In order to confirm the effects of lower values of $\sigma_8$ on the CMB and verify the observational viability of the possible solution to the $\sigma_8$ tension discussed in Sec.~\ref{sec:sigma8andTdg0}, we will next explore the best fit with low $\sigma_8$ and low $T_{{\rm d}\gamma}^0$ from the BBN-consistent model without inclusion of the SH0ES data to determine whether it does indeed provide a good fit to the \emph{Planck} CMB data.

\section{Observational viability of region of low {\boldmath$\sigma_8$ and low $T_{{\rm d}\gamma}^0$}} \label{sec:lowsigma8}

As an assessment of the observational viability of this region of low $\sigma_8$ and low $T_{{\rm d}\gamma}^0$ and as an exploration of the quality of the fit to current data provided by models in this region, we assess the best-fit BBN-consistent model with $T_{{\rm d}\gamma}^0 < 0.75$, $f_{\textrm{adm}} > 0.09$, and fixed $N_\nu = 3.044$. We chose these bounds to ensure that we selected a model with low $\sigma_8$ and to ensure that the chosen $f_{\textrm{adm}}$ is both comparable to the value of $f_{\textrm{adm}}$ analyzed in Sec.~\ref{sec:modelspace} and is high enough to have detectable effects. We chose to fix $N_\nu$ to its standard model value of 3.044 \cite{Bennett:2020zkv,Akita_2020,Froustey:2020mcq} in order to reduce the amount of beyond-$\Lambda$CDM physics introduced in this assessment. The model parameters for this best-fit model are described in the Fig.~\ref{fig:lowsigma8pwrspectra} caption.

Fig.~\ref{fig:lowsigma8pwrspectra} compares the $\Lambda$CDM TT, TE, EE, lensing, and linear matter power spectra with the corresponding power spectra for this best-fit model, as well as with current CMB and lensing data. We see that the ADM and \lcdm\ TT, TE, and EE spectra are extremely similar. The lensing power spectra are quite similar as well. In all four cases, both models appear to provide good fits to current data. The \lcdm\ and ADM matter power spectra are very similar at low $k$, with the expected differences (see Sec.~\ref{sec:modelspace}) appearing at high $k$. 

We do not show a comparison with inferences of $P(k)$ from observations. An accurate comparison here would require calculation of nonlinear corrections to the ADM model, a calculation that we have kept beyond the scope of this work. Recently, the authors of Refs.~\cite{amon2022non, preston2023non} found that the mild tension between $S_8 \equiv \sigma_8 \sqrt{\Omega_{\rm m}/0.3}$ values inferred from weak lensing shear measurements and those from \emph{Planck} \lcdm\ cosmology can be reconciled if the nonlinear matter power is more suppressed than hydrodynamical simulations predict. The scale at which this suppression emerges is similar to the scale where we start to see deviations from \lcdm\ in the linear ADM $P(k)$. ADM could provide a possible mechanism for this additional suppression.

We now present a more quantitative comparison with the calculation of the difference in $\chi^2$ values between the best-fit \lcdm\ model and our best-fit ADM model with low $\sigma_8$ and low $T_{{\rm d}\gamma}^0$. The $\chi^2$ for the ADM model is $\sim 1.71$ higher than that of the best-fit $\Lambda$CDM model. A slightly better fit with $\Delta \chi^2 \approx 1.49$ can be found if $N_\nu$ is allowed to vary, with a slightly lower $N_\nu$. However, in this case, the $\chi^2$ reduction is less than the decrease in the number of degrees of freedom, so freeing $N_\nu$ does not improve the quality of fit by more than one would expect from noise fitting. Note that, while the $\sigma_8$ value of $0.7869$ for the best-fit model with fixed $N_\nu$ is higher than the KiDS-1000 central value, it is less than their 68\% confidence upper limit of 0.79 \cite{heymans_kids-1000_2021}. Since this region of high $f_{\rm adm}$ and low $T_{{\rm d}\gamma}^0$ provides an acceptable fit to current CMB data, with a value of $\sigma_8$ in agreement with low-redshift measurements, we see that it contains possible solutions to the $\sigma_8$ tension.

Our claim that the ADM model can improve concordance with cosmic shear measurements is substantiated by the fact that it can accommodate a lower $\sigma_8$ consistent with the values inferred from these measurements. However, this claim should be verified via an analysis that compares with the data at a lower level of reduction; e.g., with the shear-shear correlation functions they determine. We believe such a study is highly motivated by the analysis we present here.

\section{Forecasting the sensitivity of future experiments to atomic dark matter} \label{sec:forecasts}

The low $\sigma_8$, low $T_{{\rm d}\gamma}^0$, high $f_{\rm adm}$ region of parameter space that is consistent with current data raises the question of its detectability with future measurements. We investigate this question for the ongoing surveys of the South Pole Telescope with its third focal plane detector array (SPT-3G), a survey to start in 2024 (the Simons Observatory or SO), and one planned to start near the end of this decade, CMB-S4 \cite{CMB-S4:2016ple, decadal_survey_on_astronomy_and_astrophysics_2020_astro2020_pathways_2021}.

\subsection{Experimental setups}
\label{sec:experiments}

\begin{figure*}
    \centering
    \includegraphics[width=0.8\textwidth]{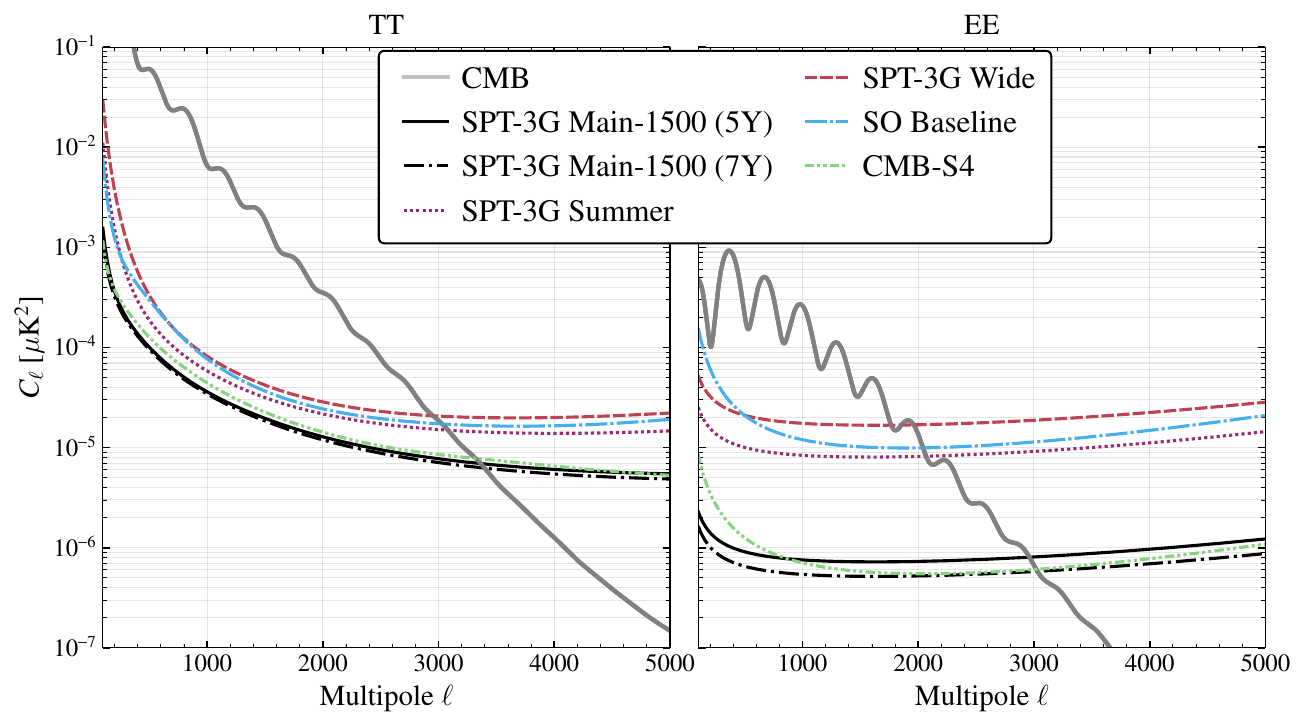}
    \caption{The total noise and extragalactic foreground residual power $N_{\ell}^{\rm MV-ILC}$ expected after the optimal internal linear combination of data from different frequency bands for different experiments considered in this work.}
    \label{fig:residual_ilc_curves}
\end{figure*}

\begin{table*}   
    \def\arraystretch{1.3}
     \begin{tabular}{c || c | c | c | c | c | c | c | c }
\multirow{2}{*}{\hspace{0.5cm}Survey\hspace{0.5cm}} & \multirow{2}{*}{\hspace{0.5cm}$\fsky$\hspace{0.5cm}} & \multirow{2}{*}{\hspace{0.5cm} Survey period \hspace{0.5cm}} & \multicolumn{6}{c}{Beam $\theta_{\rm FWHM}$ in arcminutes ($\Delta_{T}$ in $\ukarcmin$)} \\
\cline{4-9}
& & & 30 GHz & 40 GHz & 90 GHz & 150 GHz & 220 GHz & 285 GHz\\\hline
    \hline
    \mainfield{} (5Y) & 0.036 & 2019-23 & \multirow{4}{*}{-} & \multirow{4}{*}{-} & 1.7 (3) & 1.2 (2.2) & 1 (8.8) & \multirow{4}{*}{-}\\
    \mainfield{} (7Y) & 0.036  & 2019-23, 2025-26 &  &  & 1.7 (2.5) & 1.2 (1.85) & 1 (7.4) & \\
    \summerfield{} & 0.064 &  2019-23 & & & 1.7 (8.5) & 1.2 (9.0) & 1 (31) & \\
    SPT-3G Wide & 0.145 & 2024 &  &  & 1.7 (14) & 1.2 (12) & 1 (42) & \\\hline\hline
    SO Baseline & 0.4 & 2024-29 & 7.4 (71) & 5.1 (36) & 2.2 (8) & 1.4 (10) & 1 (22) & 0.9 (54)\\\hline\hline
    CMB-S4 & 0.57 & 2029-36 & 7.4 (21.3) & 5.1 (11.7) & 2.2 (1.9) & 1.4 (2.1) & 1 (6.9) & 0.9 (16.9)\\\hline\hline
     \end{tabular}   
     \caption{Band-dependent beam and white noise levels for different experiments considered in this work. The three \sptnew{} surveys: \mainfield, \summerfield, and SPT-3G Wide are surveys of non-overlapping sky patches. \sptsummerwinter{} is the combination of \mainfield{} (5Y) and \summerfield; and \sptextfield{} is the combination of three \sptnew{} surveys: \mainfield{} (7Y), \summerfield, and SPT-3G Wide. The noise in the polarization maps are assumed to be $\sqrt{2}\Delta_{T}$.}
\label{tab:exp_specs}
\end{table*}

\begin{table}
    \def\arraystretch{1.3}
    \centering
    \begin{tabular}{c||c|c||c|c}
    \multirow{2}{*}{Band [GHz]} & \multicolumn{2}{c||}{\sptnew} & \multicolumn{2}{c}{CMB-S4}\\\cline{2-5}
    & T & P & T & P \\\hline
    30 & \multirow{2}{*}{-} & \multirow{2}{*}{-} & 415, 3.5 & \multirow{7}{*}{700, 1.4} \\\cline{1-1}\cline{4-4}
    40 & & & 391, 3.5 & \\\cline{1-4}
    90 & 1200, 3.0 & \multirow{3}{*}{300, -1} & 1200, 4.2 & \\\cline{1-2}\cline{4-4}
    150 & 2200, 4.0 & & 1900, 4.1 & \\\cline{1-2}\cline{4-4}
    220 & 2100, 3.9 & & 2100, 4.1 & \\\cline{2-4}
    285 & - & - &  2100, 3.9 & \\\hline
    \end{tabular}
    \caption{Atmospheric $1/f$ noise specifications ($\ell_{\rm knee}, \alpha_{\rm knee}$ in Eq.~\eqref{eq:noise_power}) for \sptnew{} and CMB-S4 experiments. For SO Baseline, we simply adopt the procedure outlined in Refs.~\cite{ade_simons_2019, raghunathan22b}.}
\label{tab:atmnoise_specs}
\end{table}

We consider a few current and future CMB surveys: \sptnew{} \cite{sobrin_design_2022}, SO \cite{ade_simons_2019}, and CMB-S4 \cite{CMB-S4:2022ght}. For \sptnew, we perform forecasts for two setups: (A) \sptsummerwinter{} covering 4000 \sqdeg{} and (B) \sptextfield{} covering 10,000 \sqdeg. SPT-3G Ext-4k is the ongoing SPT survey, a combination of two fields (`Main-1500' and `Summer') with different noise levels with a combined coverage slightly greater than 4000 \sqdeg{}. \sptextfield{} is the combination of \sptsummerwinter{} with an additional 6000 \sqdeg{} survey expected to be carried out in the 2024 Austral Winter. 

Table \ref{tab:exp_specs} lists the experimental beam and the white noise levels for all the surveys. Other than the instrumental white noise, the maps also contain atmospheric noise parameterized using $\ell_{\text{knee}}$ and $\alpha_{\text{knee}}$ with values given in Table \ref{tab:atmnoise_specs} for \sptnew{} and CMB-S4. 
For SO Baseline, we model the atmospheric noise using the procedure described in Ref.~\cite{raghunathan22b} and adopt the values of the parameters of this noise model as given in Ref.~\cite{ade_simons_2019}.

The total noise power spectrum in each frequency band is modeled as
\begin{equation}
 N_\ell =  \Delta_{T}^2 \left[1 + \left(\frac{\ell}{\ell_{\text{knee}}}\right)^{\alpha_{\text{knee}}}\right] B_{\ell}^{-2},
\label{eq:noise_power}
\end{equation}
where $\Delta_{T}$ represents the white noise level for a given band in $\ukarcmin$; $\ell_{\text{knee}}$ and $\alpha_{\text{knee}}$ are used to parameterize the atmospheric noise; and $B_{\ell}$ is the experimental beam window function $B_{\ell} = e^{-\ell^{2}\theta_{\rm FWHM}^2/16{\rm ln 2}}$ with $\theta_{\rm FWHM}$ listed in Table \ref{tab:exp_specs}. 

\subsection{Foregrounds and internal linear combination}
\label{sec:foregrounds_ilc}

Other than the experimental noise described above, the total variance in the maps also receives contributions from astrophysical and galactic foreground emission. 

Extragalactic sources include radio galaxies (RG), dusty star-forming galaxies (DG), and kinematic and thermal Sunyaev-Zel{'}dovich (kSZ and tSZ) signals. We model these based on the SPT measurements \cite{reichardt21}. For more details of the modeling of these temperature foregrounds, we refer the reader to Refs.~\cite{raghunathan22b, raghunathan23}. For polarization, we assume the 2\% polarization fraction for RG and simply scale the temperature power spectrum of $C_{\ell}^{\rm RG}$ accordingly \cite{datta18, gupta19}. Other extragalactic foregrounds are assumed to be unpolarized. 

Given that surveys like CMB-S4 plan to scan larger regions of sky, it is crucial to consider the impact of galactic foreground signals as well. Important galactic foregrounds include galactic dust and synchrotron signals. While the original CMB-S4 footprint is $\sim 67\%$, we ignore the regions ($-10^{\circ} \le b_{\rm gal} \le 10^{\circ}$) that are strongly contaminated by our Galaxy and only consider 57\% for the forecasting purposes here. The \sptnew{} Ext-4k, \sptnew{} Ext-10k, and SO Baseline footprints, with sky fractions of $\sim 10\%, \ \sim 25\%$, and $\sim 40\%$ respectively, avoid regions with strong galactic foregrounds. We therefore ignore galactic foregrounds for the baseline \sptnew{} and SO forecasts, as well as for the CMB-S4 forecasts after the galactic plane cut. However, as a sanity check, we assess the impact of galactic foregrounds using pySM \citep{thorne17} simulations. This is described in Sec.~\ref{sec:forecastresults}.

To reduce the overall impact of noise and foregrounds, we use the internal linear combination (ILC) technique \cite{cardoso08, planck14_smica} to optimally combine data from multiple frequency bands to produce the minimum variance (MV-ILC) CMB map as
\begin{equation}
	S_{\ell m} = \sum_{i=1}^{N_\mathrm{bands}} w_\ell^i M_{\ell m}^i\, ,
\end{equation}
where $M_{\ell m}^i$ is the spherical harmonic transform of the map from the $i$th frequency band and the multipole-dependent weights $w_\ell^i$ are tuned in order to minimize the overall variance from noise and foregrounds. The weights are derived as 
\begin{equation}
	w_\ell = \frac{\clinv A_{\rm s}}{A_{\rm s}^\dagger \clinv A_{\rm s}}\, , 	\label{eq:ilc_weights_mv}
\end{equation}
where $\clcov$ is a $N_\mathrm{bands} \times N_\mathrm{bands}$ matrix containing the covariance of foregrounds and beam-deconvolved noise across different frequencies at a given multipole $\ell$; $A_{\rm s} = [1, 1, ..., 1]$ is the frequency response of the CMB in different bands, and it is a $N_\mathrm{bands} \times 1$ vector.
The sum of residual noise and foregrounds power in the MV-ILC map is given as 
\begin{equation}
N_{\ell}^{\rm MV-ILC} = \dfrac{1}{{A_{\rm s}}^{T} \clinv {A_{\rm s}}}\, ,
\end{equation}
which we use for forecasting in the subsequent sections. In Fig.~\ref{fig:residual_ilc_curves}, we present $N_{\ell}^{\rm MV-ILC}$ for all the experiments considered in this work. 

To include CMB lensing information, we feed the MV-ILC residuals $N_{\ell}^{\rm MV-ILC}$ into a lensing quadratic estimator (QE) \cite{hu02, okamoto03} and compute the resultant lensing noise curves $N_\ell^{\phi\phi}$. We set $\ell_{\rm max} = 3500$ for temperature and $\ell_{\rm max} = 4000$ for polarization during lensing reconstruction. The former choice is to mitigate the impact of mismodeling the small-scale foregrounds in the temperature maps. For lensing reconstruction from a more optimistic version of CMB-S4, we increase to $\ell_{\rm max} = 5000$ for both temperature and polarization.

\subsection{Forecasting methods} \label{sec:forecastmethods}

We forecast expected errors on ADM parameters from measurements of CMB intensity and linear polarization maps as follows. First, we calculate the Fisher information matrix via
\begin{equation}
    F_{ij} = \displaystyle\sum_{\ell} \left(\frac{\partial C_\ell^\nu}{\partial \alpha_i}\right)^T \cdot \left(\Sigma_\ell^{-1}\right)_{\nu \nu'} \cdot \frac{\partial C_\ell^{\nu'}}{\partial \alpha_j},
    \label{eq:F_ij}
\end{equation}
where $\nu = \{TT, EE, TE, \phi \phi\}$; the $\alpha_{i}$ are the model parameters; and
\begin{multline}
    \Sigma_\ell = \tfrac{2}{\left(2\ell+1\right)f_{\rm sky}} \times \\ {\tiny \begin{pmatrix}
(\tilde C_\ell^{TT})^2 & (C_\ell^{TE})^2 & \tilde C_\ell^{TT} C_\ell^{TE} & (C_\ell^{T\phi})^2\\
(C_\ell^{TE})^2 & (\tilde C_\ell^{EE})^2 & \tilde C_\ell^{EE} C_\ell^{TE} & (C_\ell^{E\phi})^2\\
\tilde C_\ell^{TT} C_\ell^{TE} & \tilde C_\ell^{EE} C_\ell^{TE} & \frac{1}{2}\left[(C_\ell^{TE})^2 + \tilde C_\ell^{TT}\tilde C_\ell^{EE}\right] & C_\ell^{T\phi}C_\ell^{E\phi}\\
(C_\ell^{T\phi})^2 & (C_\ell^{E\phi})^2 & C_\ell^{T\phi}C_\ell^{E\phi} & (\tilde C_\ell^{\phi\phi})^2 \end{pmatrix}},
    \label{Sigma_l}
\end{multline}
where
\begin{equation}
    \tilde C_\ell^{XX} = C_\ell^{XX} + N_\ell^{XX},
    \label{eq:Clnoisesub}
\end{equation}
$XX = \{TT, EE, \phi\phi\}$, $N_{\ell}^{XX}$ is the residual foregrounds and noise spectrum for the MV-ILC map $N_{\ell}^{\rm MV-ILC}$ described in Sec.~\ref{sec:foregrounds_ilc}, 
and $f_{\textrm{sky}}$ is the fraction of the sky observed. We then calculate the parameter covariance matrix as $C = F^{-1}$. Our Eq.~\eqref{Sigma_l} can be derived from Eqs.~(8.20) and (8.21) of Ref.~\cite{CMB-S4:2016ple}. For all experiments, we include $100 \leq \ell \leq 3500$ for TT, $100 \leq \ell \leq 4000$ for EE/TE, and $30 \leq L \leq 3500$ for $\phi\phi$ to match the $\ell$ ranges used for lensing reconstruction. As mentioned earlier in Sec.~\ref{sec:foregrounds_ilc}, we ignore small-scale ($\ell > 3500$) temperature information due to the difficulties in modeling the foregrounds. Given that CMB-S4 is a expected to operate in the next decade, we also consider an optimistic case by setting $\ell_{\rm max} = 5000$, assuming we will have better foreground mitigation strategies by then. To be specific, we use $100 \leq \ell \leq 5000$ for TT/EE/TE and $30 \leq L \leq 5000$ for $\phi\phi$.

\begin{table*}
\centering
\begin{tabular}{cccccccc}
    \hline \hline 
    \multicolumn{1}{c}{} & \multicolumn{1}{c}{} & \multicolumn{4}{c}{Forecasted standard error ($\sigma$)} \\
    Parameter & \ Fiducial value \ & \ SPT-3G Ext-4k \ & \ SPT-3G Ext-10k \ & \ SO Baseline \ & \ CMB-S4 \ \\ \cmidrule(r){1-1} \cmidrule(rl){2-2} \cmidrule(l){3-6}

    $f_{\textrm{adm}}$ & $0.102$ & 0.032 & 0.025 & 0.025 & 0.009 \\
    $B_{\rm d}/T_{{\rm d}\gamma}^0$ (eV/K) & $15$ & 21 & 14 & 11 & 8 \\
    $T_{{\rm d}\gamma}^0$ (K) & $0.50$ & 0.14 & 0.10 & 0.08 & 0.05 \\
    $N_{\textrm{eff}}$ & $3.05$ & 0.12 & 0.09 & 0.08 & 0.04 \\

    \cmidrule(r){1-1} \cmidrule(rl){2-2} \cmidrule(l){3-6}
    $f_{\textrm{adm}}$ & $0.030$ & 0.023 & 0.016 & 0.014 & 0.008 \\
    $B_{\rm d}/T_{{\rm d}\gamma}^0$ (eV/K) & $13.0$ & 5.3 & 3.6 & 3.0 & 2.0 \\
    $T_{{\rm d}\gamma}^0$ (K) & $1.39$ & 0.36 & 0.26 & 0.22 & 0.13 \\
    $N_{\textrm{eff}}$ & $3.25$ & 0.16 & 0.12 & 0.10 & 0.05 \\
    \hline \hline
\end{tabular}
\caption{Forecasted constraints on SPT-3G Ext-4k, SPT-3G Ext-10k, the Simons Observatory (SO) Baseline, and CMB-S4. 
Top: Forecasted constraints on the best-fit BBN-consistent model with $T_{{\rm d}\gamma}^0 < 0.75$, $f_{\textrm{adm}} > 0.09$, and fixed $N_\nu = 3.044$, as explored in Sec.~\ref{sec:lowsigma8} and Fig.~\ref{fig:lowsigma8pwrspectra}. In this model, $H_0 = 67.52$ $\textrm{km} \; \textrm{s}^{-1} \; \textrm{Mpc}^{-1}$, $\Omega_{\rm b} h^2 = 0.02241$, $\Omega_{\rm c} h^2 = 0.1076$, $\tau = 0.05768$, $A_{\rm s} = 2.113 \times 10^{-9}$, $n_{\rm s} = 0.9667$, and $\sigma_8 = 0.7869$.
Bottom: Forecasted constraints on the best-fit BBN-consistent model with fixed $f_{\textrm{adm}} = 0.03$. In this model, $H_0 = 68.58$ $\textrm{km} \; \textrm{s}^{-1} \; \textrm{Mpc}^{-1}$, $\Omega_{\rm b} h^2 = 0.02266$, $\Omega_{\rm c} h^2 = 0.1213$, $\tau = 0.05721$, $A_{\rm s} = 2.1175 \times 10^{-9}$, $n_{\rm s} = 0.9714$, and $\sigma_8 = 0.7974$.}
\label{tab:forecasts}
\end{table*}

\begin{figure*}
    \centering
    \begin{subfigure}[b]{\columnwidth}
        \centering
        \includegraphics[width=\columnwidth]{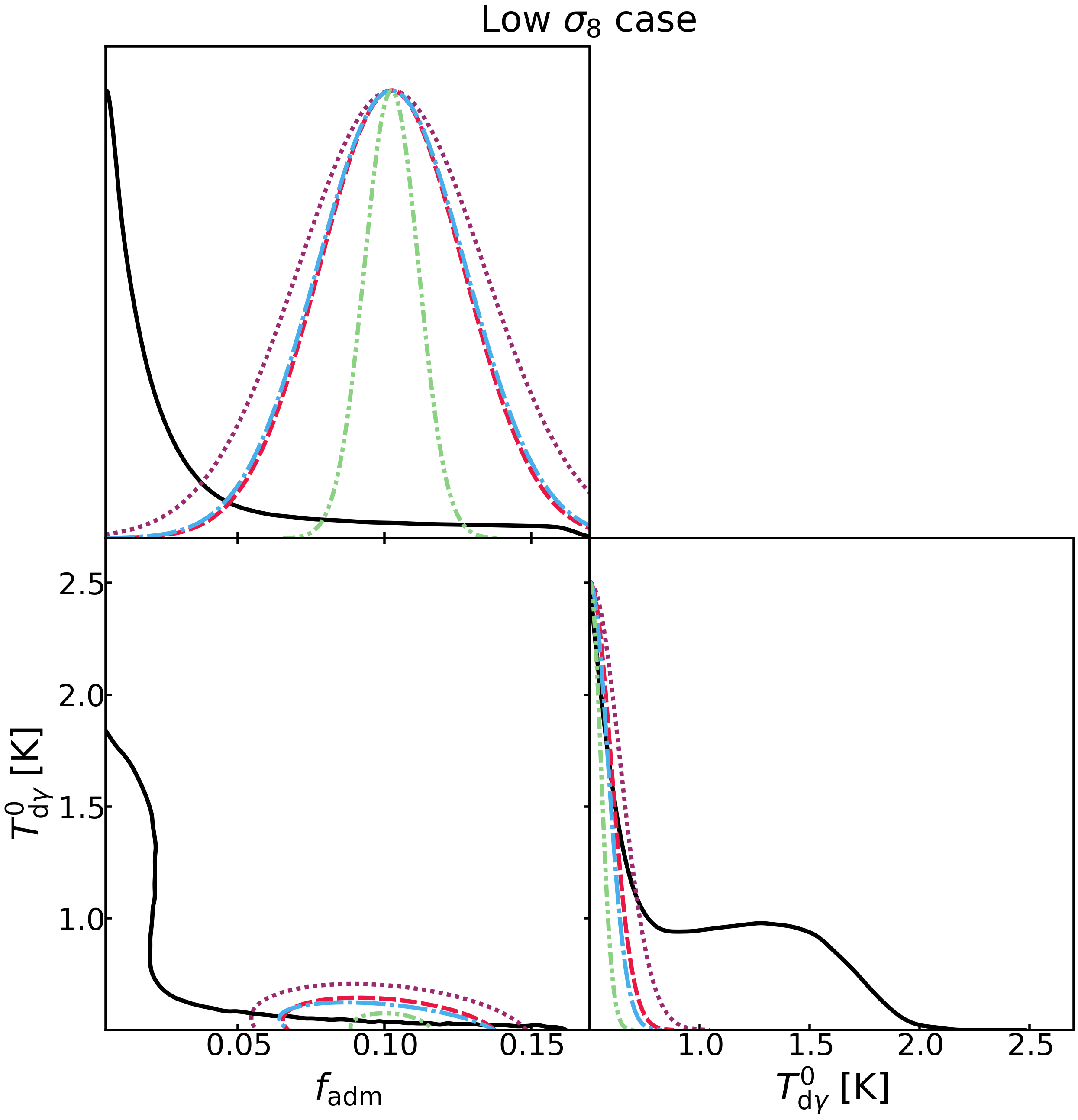}
    \end{subfigure}
    \hfill
    \begin{subfigure}[b]{\columnwidth}
        \centering
        \includegraphics[width=\columnwidth]{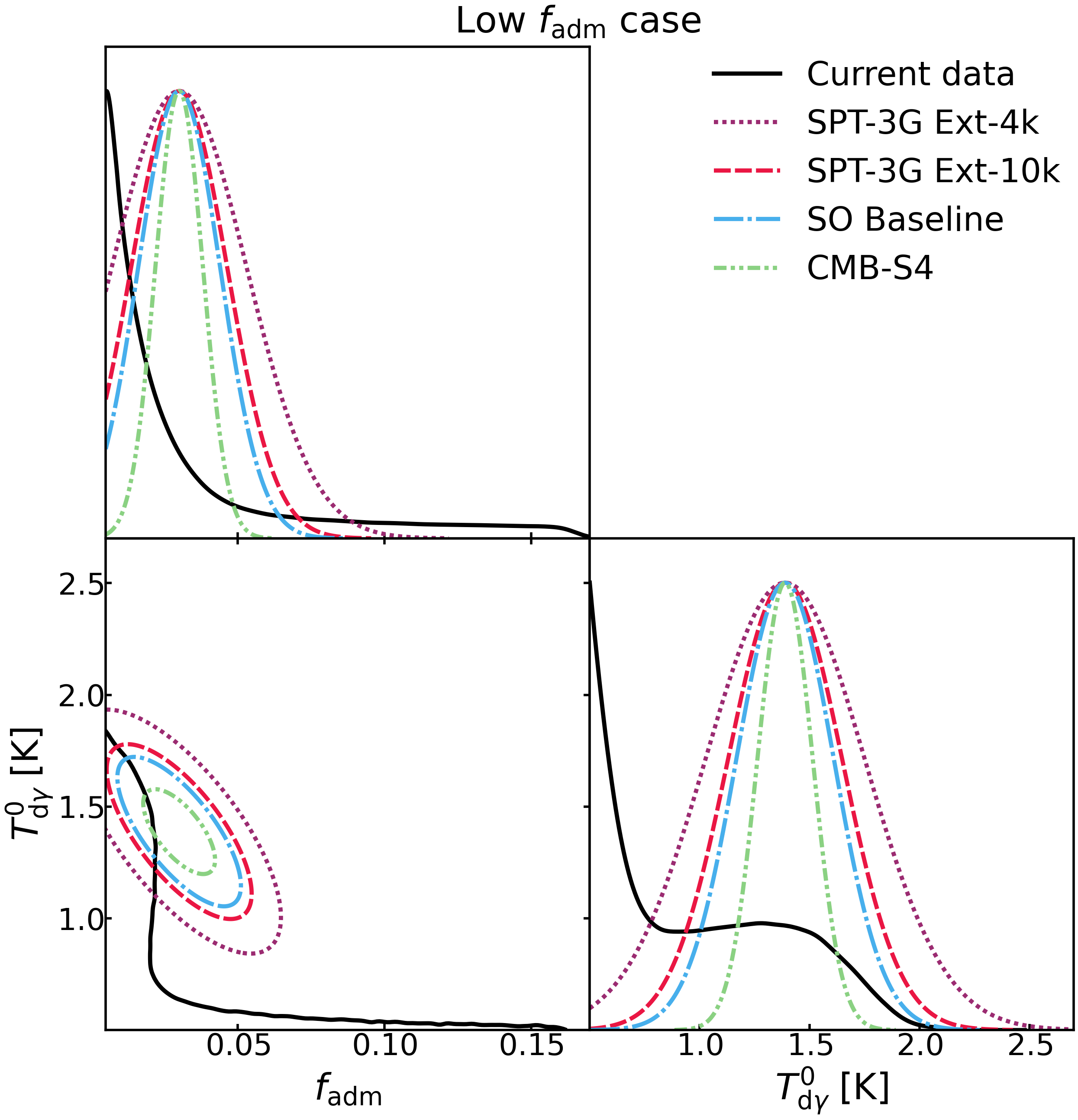}
    \end{subfigure}
    \hfill
    \caption{1- and 2-dimensional marginal posterior probability densities for $f_{\textrm{adm}}$ and $T_{{\rm d}\gamma}^0$ for the BBN-consistent model without the inclusion of the SH0ES data, as well as the forecasted 1- and 2-dimensional posteriors for $f_{\textrm{adm}}$ and $T_{{\rm d}\gamma}^0$ for SPT-3G Ext-4k, SPT-3G Ext-10k, SO Baseline, and CMB-S4. The (arbitrarily normalized) vertical axes of the diagonal indicate probability. The contours on the 2-dimensional posteriors enclose the 68\% credible region. Left: Forecasted constraints on the low $\sigma_8$ model outlined in the top part of Table \ref{tab:forecasts}. Right: Forecasted constraints on the low $f_{\rm adm}$ model outlined in the bottom part of Table \ref{tab:forecasts}.}
    \label{fig:twomodelsforecasts}
\end{figure*}

This procedure is appropriate for homogeneous coverage, when correlations across multipole moments and types of spectra due to gravitational lensing can be neglected, and when the posterior probability distribution of the model parameters is close to Gaussian. The impacts of lensing-induced correlations increase with decreasing noise levels. Even for CMB-S4, these effects tend to cause at most 20\% changes to parameter standard errors \cite{Hotinli:2021umk}. Non-Gaussianity of the parameter posterior distribution is a concern we address below.

\subsection{Forecasted sensitivity of future experiments} \label{sec:forecastresults}

In order to forecast the sensitivity of upcoming CMB experiments to the low $\sigma_8$ and low $T_{{\rm d}\gamma}^0$ region of parameter space discussed in Sec.~\ref{sec:sigma8andTdg0} and Sec.~\ref{sec:lowsigma8}, we will forecast the sensitivity of SPT-3G, SO, and CMB-S4 to the best-fit BBN-consistent model with $T_{{\rm d}\gamma}^0 < 0.75$, $f_{\textrm{adm}} > 0.09$, and fixed $N_\nu = 3.044$ that we explored in Sec.~\ref{sec:lowsigma8}, hereafter referred to as the low $\sigma_8$ model. As a second point of reference, we will also forecast the sensitivity of SPT-3G, SO, and CMB-S4 to the best-fit BBN-consistent model with fixed $f_{\textrm{adm}} = 0.03$, hereafter referred to as the low $f_{\rm adm}$ model. The fiducial models used for forecasts, as well as one-dimensional marginal errors in ADM parameters and $N_{\textrm{eff}}$, are outlined in Table \ref{tab:forecasts}. Forecast results are also shown in Fig.~\ref{fig:twomodelsforecasts}.

Before looking at the ADM parameter constraints, we point out that the forecasted constraints on $N_{\rm eff}$ degrade some from their \lcdm\ + $N_{\rm eff}$ expected values due to the additional degrees of freedom. For example, the CMB-S4 forecasted $\sigma(N_{\textrm{eff}}) = 0.04$ for the low $\sigma_8$ fiducial model is somewhat larger than the forecasted $\sigma(N_{\textrm{eff}}) = 0.0327$ for a fiducial model with no ADM \cite{cmbs4dsr19}. If we include up to $\ell = 5000$ for TT/EE/TE/$\phi\phi$ and for lensing reconstruction, as done in \cite{raghunathan23}, we also find $\sigma(N_{\textrm{eff}}) = 0.04$. Presumably this mild degradation to 0.04 from 0.0327 is due to some degree of degeneracy that emerges between $N_{\rm eff}$ and the other cosmological parameters, with the addition of the three ADM parameters. We have checked that when we instead eliminate the ADM, using an unmodified version of CAMB, we recover $\sigma(N_{\rm eff}) = 0.033$. 

The key question we wish to address here is whether CMB-S4 will be able to detect the influence of ADM if the cause of the $\sigma_8$ tension is indeed due to a low-temperature ADM. We are thus most interested in the constraints on the low $\sigma_8$ fiducial model -- the one with $f_{\rm adm} = 0.102$ used for the left panel of Fig.~\ref{fig:twomodelsforecasts}. As can be seen in Table \ref{tab:forecasts}, the expected error on $f_{\rm adm}$ in this case is 0.009, allowing for a greater than $11\sigma$ detection. If we include up to $\ell=5000$ for TT/EE/TE/$\phi\phi$ and lensing reconstruction, which would require successfully cleaning foregrounds out to higher $\ell$, the expected error nearly halves to 0.005, which would allow for a greater than $20\sigma$ detection. 

The posteriors given current data, as one can see in Figs.~\ref{fig:2dposteriors} and \ref{fig:twomodelsforecasts}, are highly non-Gaussian. The forecasting methods used assume Gaussianity, so the non-Gaussian posterior is a source of concern. However, we expect that as the data become more constraining, the posterior will become more Gaussian. We see evidence of this from the forecasts for our second fiducial model, which has a much lower value of $f_{\rm adm}$. The error on $f_{\rm adm}$ assuming this fiducial model is 0.008, only slightly different from the higher $f_{\rm adm}$ (low $\sigma_8$) case. The marginal posterior for $f_{\rm adm}$ given CMB-S4 data thus appears to be fairly Gaussian, enough so that an $11\sigma$ detection would almost certainly be a detection of very high significance. 

Experiments prior to CMB-S4 also have a chance of seeing some, perhaps only weak, evidence in favor of non-zero $f_{\rm adm}$. SPT-3G Ext-10k and SO Baseline, given the low $\sigma_8$ fiducial model, can nominally be expected to make $4\sigma$ detections of non-zero $f_{\rm adm}$. However, one can see greater fiducial model sensitivity to the error on $f_{\rm adm}$ for SPT-3G Ext-4k, SPT-3G Ext-10k, and SO Baseline, so these constraints are both weaker than those from CMB-S4 and less certain. A clearer picture of expectations in these cases would require calculation of posterior distributions from simulated data.

We also note that these forecasts indicate that some fairly precise measurements of $T_{{\rm d}\gamma}^0$ may be possible with CMB-S4 data in either fiducial model case; formally, these are $\sim 10\sigma$ constraints. In the low $T_{{\rm d}\gamma}^0$ (low $\sigma_8$) case, one should be able to both bound the temperature from below, and have clear evidence that the temperature is well below that of the CMB or the cosmic neutrino background. The temperature constraint would be a valuable clue, in either case, about a possible thermal origin for the dark photon relic.

We have also plotted contours from our forecasts and for current data for our two most well-constrained ADM parameters, $f_{\rm adm}$ and $T_{{\rm d}\gamma}^0$, in Fig.~\ref{fig:twomodelsforecasts}. 
One can visually see here the qualitative improvement over constraints from current data expected very soon from the SPT-3G Ext-4k survey, and a continuing progression through to the CMB-S4 constraints. 

As mentioned in Sec.~\ref{sec:foregrounds_ilc}, we also check the impact of galactic foregrounds on our constraints. We generate galactic dust and synchrotron signals using publicly available pySM \citep{thorne17} simulations, which are based on {\it Planck} sky model code \citep{delabrouille13}. We calculate the dust and synchrotron foreground power in the SPT and CMB-S4 footprints, and include them along with other components in the ILC step. With the inclusion of galactic foregrounds, we find that the above constraints weaken only marginally by $\lesssim 5\%$.

It is important to note that all forecasts in this paper are based on linear theory predictions. However, there exist significant nonlinear corrections to the results outlined in this section, meaning that future work on nonlinear predictions is needed to refine these forecasts, as well as for the comparison of theoretical matter power spectra to data mentioned in Sec.~\ref{sec:lowsigma8}.

Despite a caveat regarding the varying sensitivity of future experiments to different fiducial ADM models, ultimately, these forecasts demonstrate the promise for the ADM model in general, and the low $\sigma_8$ and low $T_{{\rm d}\gamma}^0$ region of parameter space in particular, to be tightly constrained with upcoming CMB experiments. These tighter constraints include the possibility of either confirming or excluding the low $\sigma_8$ and low $T_{{\rm d}\gamma}^0$ region of parameter space, meaning that, if this region is the solution to the $\sigma_8$ tension, its effects should be detectable with future experiments.

\section{Conclusions} \label{sec:conclusions}

We investigated the observational consequences of an extension to \lcdm\ to include both ADM and the possibility of a non-standard energy density in free-streaming massless particles. The ADM consists of dark protons and dark electrons interacting via a dark electromagnetism, and hence the model includes a thermal background of dark photons. Like in the visible sector, these undergo a transition from a tightly coupled plasma to dark hydrogen plus free-streaming dark photons. Since the dark electrons, protons, and hydrogen are all non-relativistic during the epochs of interest we consider them as contributions to the total density of dark matter, while the massless dark photons contribute to the total radiation density. The free parameters of the model we took to be the fraction of dark matter that is atomic $f_{\rm adm}$, the temperature of dark photons today $T_{{\rm d}\gamma}^0$, and the binding energy of dark hydrogen $B_{\rm d}$, in addition to the standard \lcdm\ parameters and $N_{\rm eff}$.

We examined the impact of ADM on CMB and matter power spectra in the ``cool" regime (with $T_{{\rm d}\gamma}^0$ in the 0.6 K to 1 K range) in Sec.~\ref{sec:modelspace}. In this regime, the main influence on the CMB temperature and polarization spectra is via gravitational lensing, due to the impact on the matter power spectrum $P(k)$. We saw no change from $P(k)$ in our baseline \lcdm\ model for modes entering the horizon after dark recombination, and a nearly uniform suppression of power for modes that enter during radiation domination and while $c_{\rm s,d}^2$ has not yet dropped much below its maximum value of 1/3. The suppression of power is due to the pressure support received by the fraction of the dark matter that is ADM. This pressure support slows down the gravitationally driven growth of both the ADM and CDM perturbations, as the ADM does not contribute as much as it would otherwise to gravitational potential gradients. Such suppression of power is a general feature of models in which part of the dark matter interacts with dark radiation, which others have also noted might address the $\sigma_8$ tension \cite[e.g.][]{Buen-Abad:2015ova,Chacko:2016kgg,Buen-Abad:2017gxg,Joseph:2022jsf,Buen-Abad:2022kgf,Rubira:2022xhb,Buen-Abad:2023uva,Aloni:2021eaq,Gariazzo:2023hch,Schoneberg:2023rnx,bansal_precision_2022}.

We also looked at the dependence of these effects on variations in the dark photon temperature in Sec.~\ref{sec:changeTd0}, and on variations in the dark hydrogen binding energy in Sec.~\ref{sec:changeBd}. Because of the primary role of CMB lensing, we also focused here on $P(k)$. Variation in $B_{\rm d}$ (at fixed $T_{{\rm d}\gamma}^0$) changes when dark recombination occurs, impacting the duration of pressure support for the ADM and, therefore, the amount of power suppression. Varying $T_{{\rm d}\gamma}^0$ also changed the amplitude of dark photon pressure. We looked at the dependence of the square of the sound speed $c_{\rm s,d}^2$ of the dark photon-ADM plasma on dark photon temperature since pressure gradients are proportional to this quantity. No matter the dark photon temperature, at sufficiently early times, $c_{\rm s,d}^2$ reaches its maximal value of 1/3. Modes that enter the horizon after significant reduction of this quantity do not suffer the suppression in growth due to pressure support. Reducing $T_{{\rm d}\gamma}^0$ thus moves the onset of $P(k)$ suppression to higher $k$.

We then constrained the ADM model using current CMB \cite{planck_collaboration_planck_2020} and BAO \cite{beutler_6df_2011,ross_clustering_2015,alam_clustering_2017} data in Sec.~\ref{sec:constraints} by considering four different model spaces: both with and without BBN constraints on helium abundance, and with and without incorporating the $H_0$ measurement from SH0ES \cite{riess_comprehensive_2022}. In Sec.~\ref{sec:scalingsym}, we saw in elongated probability density contours evidence of a quasi-degeneracy between $f_{\textrm{adm}}$, $T_{{\rm d}\gamma}^0$, and $H_0$. We pointed out that this is expected since this model space allows for a scaling transformation that is nearly the FFAT-scaling transformation, with associated symmetry, introduced in Ref.~\cite{cyr-racine_symmetry_2022} and discussed extensively in Ref.~\cite{ge_scaling_2022}. Likewise, this scaling transformation cannot be followed in the three cases where we included SH0ES data and/or enforced BBN consistency, and hence we do not see the elongated contours in these cases. These results highlight the explanatory power of the FFAT scaling transformation symmetry.

In Sec.~\ref{sec:sigma8andTdg0}, we presented a key finding: at low dark photon temperature, the ADM parameter space opens up significantly, allowing for higher fractions of ADM and therefore a lower $\sigma_8$. This lower $\sigma_8$ is in line with results from cosmic shear and other large-scale structure measurements \cite{heymans_kids-1000_2021,HSC:2018mrq, Hang:2020gwn, Garcia-Garcia:2021unp, DES:2021wwk, White:2021yvw, DES:2022xxr}, providing a possible solution to the $\sigma_8$ tension. We can understand the lower $\sigma_8$ physically because, as we found in Sec.~\ref{sec:modelspace}, $P(k)$ is suppressed on small scales with the presence of the pressure-supported ADM.

We subsequently verified the observational viability of the low $\sigma_8$ and low $T_{{\rm d}\gamma}^0$ region of parameter space by comparing CMB power spectra for a best-fit model in the region, to current data in Sec.~\ref{sec:lowsigma8}. We found that the best-fit model with low $T_{{\rm d}\gamma}^0$, high $f_{\rm adm}$, and $N_\nu$ fixed to its standard model value provides a good fit to current data, roughly comparable to that of \lcdm\ with $\chi^2$ only $\sim 1.71$ higher. We saw no significant improvement in quality of fit by letting $N_\nu$ vary. However, it is worth noting, as seen in Sec.~\ref{sec:H0}, that this possible solution to the $\sigma_8$ tension likely does not provide a simultaneous solution to the $\sigma_8$ and $H_0$ tensions.

This low $\sigma_8$ region of the parameter space requires a dark photon temperature of about 0.5 K. This is below the temperature expected for a boson with two degrees of freedom even if it froze out while all standard model particles were still relativistic. Such a cool component would point to different post-inflation reheating temperatures for the ADM and dark photons, relative to standard model particles, and thus have implications for the reheating process (see, e.g., Refs.~\cite{Adshead:2016xxj,Berezhiani_1996}). A non-thermal mechanism for dark-sector particle production could be a potential solution (see, e.g., Refs.~\cite{Joseph:2022jsf,Aloni:2023tff}).

Our investigation also motivates the development of nonlinear corrections to the linear ADM model predictions presented in this paper, both in order to compare ADM matter power spectra with current data and to improve forecasts, as mentioned in Secs.~\ref{sec:lowsigma8} and \ref{sec:forecastresults}. Additionally, beyond the analysis via $\sigma_8$ presented in this paper, our work highlights the need for more robust comparison to cosmic shear measurements (possibly via shear-shear correlation functions) to see the extent to which the low $\sigma_8$ region identified in Sec.~\ref{sec:sigma8andTdg0} improves concordance with these measurements, as mentioned in Sec.~\ref{sec:lowsigma8}.

In order to determine whether this possible solution to the $\sigma_8$ tension could be tested with future observations, we forecasted the sensitivity of future CMB experiments to this low $\sigma_8$ region of parameter space. We found in Sec.~\ref{sec:forecastresults} that CMB-S4 will be able to tightly constrain both the fraction of ADM and the dark photon temperature, including the possibility of detecting nonzero $f_{\rm adm}$ with high significance. We also see the promise of experiments prior to CMB-S4 -- SPT-3G and SO -- to provide steadily tighter constraints on the ADM model. Ultimately we find that, if the low $\sigma_8$ and low $T_{{\rm d}\gamma}^0$ region of parameter space is the solution to the $\sigma_8$ tension, its effects should be detectable with future CMB experiments.

\section{Acknowledgements} \label{sec:acknowledgements}

This material is based upon work initiated as part of the Physics Research Experiences for Undergraduates (REU) Program at the University of California, Davis and was supported by the National Science Foundation (NSF) under grant PHY-2150515. LK, FG, and EH were supported in part by DOE Office of Science award DE-SC0009999. F.-Y.~C.-R. is supported by the NSF under grant AST-2008696. F.-Y.~C.-R. also thanks the Robert E. Young Origins of the Universe Chair fund for its generous support. SR is supported by the Center for AstroPhysical Surveys (CAPS) at the National Center for Supercomputing Applications (NCSA), University of Illinois Urbana-Champaign. LK acknowledges the hospitality of TTK at RWTH where some of this work was completed. We thank B.~Benson and T.~Crawford for feedback on a draft and G.~Lynch for useful conversations.

\bibliography{bibliography}

\end{document}